\documentclass[sigconf]{acmart}
\AtBeginDocument{%
  }

\copyrightyear{2025}
\acmYear{2025}
\setcopyright{acmlicensed}\acmConference[DIS '25]{Designing Interactive Systems Conference}{July 5--9, 2025}{Funchal, Portugal}
\acmBooktitle{Designing Interactive Systems Conference (DIS '25), July 5--9, 2025, Funchal, Portugal}
\acmDOI{10.1145/3715336.3735707}
\acmISBN{979-8-4007-1485-6/2025/07}
\usepackage{tabularx}
\usepackage{tabularx} % 在导言区添加此包
\usepackage{multirow} % 在导言区添加此包
\usepackage{array} % 在导言区添加这个包
\usepackage{booktabs} 

\newcommand{\hh}[1]{#1}

\newcommand{\xl}[1]{#1}

\begin{document}

\title{Conversations With The Stressed Body}
\subtitle {Facilitating Stress Self-Disclosure Among Adolescent Girls Through An Embodied Approach }


\author{Xinglin Sun}
\email{xinglinsun@tongji.edu.cn}
\orcid{0009-0003-8752-202X}
\affiliation{
  \institution{College of Design and Innovation \\ Tongji University}
  \city{Shanghai}
  %\state{California}
  \country{China}}

\author{Caroline Claisse}
\email{caroline.claisse@newcastle.ac.uk}
\orcid{0000-0003-4002-6136}
\affiliation{
  \institution{Open Lab \\Newcastle University}
  \city{Newcastle upon Tyne}
  %\state{California}
  \country{United Kingdom}}

\author{Runhua Zhang}
\orcid{0000-0002-0519-5148}
\email{runhua.zhang@connect.ust.hk}
\affiliation{
  \institution{The Hong Kong University of Science and Technology}
  \city{Hong Kong SAR}
  %\state{California}
  \country{China}}

\author{Xinyu Wu}
\orcid{0009-0002-8870-8793}
\email{wuxinyu2003@tongji.edu.cn}
\affiliation{
  \institution{College of Design and Innovation \\Tongji University}
  \city{Shanghai}
  %\state{California}
  \country{China}}

\author{Jialin Yuan}
\orcid{0009-0001-1125-3383}
\email{yuanjialin@tongji.edu.cn}
\affiliation{
  \institution{College of Design and Innovation \\Tongji University}
  \city{Shanghai}
  %\state{California}
  \country{China}}

\author{Qi Wang}
\orcid{0000-0002-2688-8306}
\email{qiwangdesign@tongji.edu.cn}
\affiliation{
  \institution{College of Design and Innovation \\Tongji University}
  \city{Shanghai}
  %\state{California}
  \country{China}}

\renewcommand{\shortauthors}{Sun et al.}

\begin{abstract}
Adolescent girls face significant mental health challenges during their transition to adulthood, often experiencing heightened stress from various sources. While various interactive technologies for self-disclosure had been explored to support stress relief, little is known about how to encourage stress-related self-disclosure through an embodied approach. This study presents a co-design workshop centred on Embodied Probes—a series of artefacts and activities incorporating embodied methods and technologies. During the workshop, nine participants aged 15-18 engaged with their bodies, expressed bodily sensations through tangible means, and designed embodied prototypes tailored to their personal needs for stress perception and relief. The workshop revealed insights into somatic symptoms, sources, and coping strategies for stress among adolescent girls, as well as how embodied methods can support their stress self-disclosure. This paper contributes to the HCI community by offering design implications on leveraging embodied technologies to support self-disclosure for young women’s mental well-being.
%Adolescent girls face significant mental health challenges during during their transition to adulthood, often experiencing heightened stress from various sources. While self-disclosure has been explored as a method to support stress management, little is known about how to encourage stress-related self-disclosure through embodied approaches. This study presents a workshop centered on Embodied Probes—a series of artefacts and activities incorporating embodied techniques or technologies. During the workshop, nine participants aged 15-18 engaged with their bodies, expressed bodily sensations through tangible means, and designed embodied prototypes tailored to their personal needs for stress perception and relief. The workshop revealed insights into somatic symptoms, sources, and coping strategies for stress among adolescent girls, as well as how embodied approaches can support their stress self-disclosure. This paper contributes to HCI community by offering design implications on leveraging embodied technologies to support young women’s mental well-being through self-disclosure. 

\end{abstract}



%\keywords{Women’s mental well-being, Youth-centred, Bodily experience, Self-disclosure, Participatory Design}


\maketitle

\section{Introduction}
\hh{Stress is a pervasive challenge during adolescence, yet for girls it can be especially heightened due to unique physiological factors such as changes in body image \cite{piran_handbook_2019} and distinct social or cultural expectations on gender norms \cite{sontag_coping_2008, haraldsson_adolescent_2011}. For example, adolescent girls are often expected to be physically attractive \cite{sinton2006individual} and socially agreeable \cite{tate2022personality}.} These gender-specific expectations around appearance and behaviour often place additional burdens on adolescent girls, contributing to increased emotional vulnerability or potential mental disorder. A cross-sectional study from China shows that female students have a higher incidence of emotional disorder by 39\% compared to males \cite{wang_large-scale_2023}, while survey of adolescents across the UK showed that one in four young women aged 17-19 years suffers from major depression or anxiety, half of whom had self-harmed \cite{noauthor_mental_2017}. Though prior work in the Human-Computer Interaction (HCI) community has explored stress navigation (e.g., stress recognition, communication, and timely intervention) for adolescents as a whole \cite{rasouli_proposed_2022, jusoh_helpbot_2024, kitson_i_2024, catalan_storytelling_2024}, there is limited attention on tailoring these efforts specifically to the needs of adolescent girls.

%These standards not only reinforce narrow ideals of femininity but also increase vulnerability to self-criticism and anxiety \cite{}. Recent studies underscore the impact of these pressures:

%Adolescent girls often face heightened stress during adolescence due to various factors, such as changes in body image \cite{piran_handbook_2019}, relationship or academic pressures \cite{chyu_associations_2022}. While stress is nearly universal among adolescents in general, girls experience unique stressors influenced by biological, social, and cultural factors \cite{sontag_coping_2008, haraldsson_adolescent_2011}. these gender-specific expectations around appearance and behaviour often place additional burdens on adolescent girls , contributing to increased emotional vulnerability or potential mental disorder. 

Regarding stress navigation, self-disclosure, the deliberate sharing of personal information with others \cite{cozby_self-disclosure_1973}, has been recognised as a powerful approach. By facilitating the expression of personal thoughts and emotions, self-disclosure has demonstrated its effectiveness in reducing early symptoms of anxiety and depression, alleviating feelings of shame, and increasing the likelihood of seeking help \cite{gonsalves_systematic_2023}. Interactive technologies, such as social media \cite{andalibi_disclosure_2020} and chatbots \cite{li_tell_2023}, have further encouraged and amplified the potential of self-disclosure regarding stress. \hh{However, most existing HCI interventions supporting self-disclosure primarily rely on textual or verbal modes of expression \cite{andalibi_disclosure_2020, li_tell_2023, park_can_2020, pinch_subtleties_2024}, which may not fully address the embodied nature of stress experiences \cite{francis_embodied_2018}.}

\hh{Stress often manifests physically in the bodily sensations through muscle tension, changes in posture, increased heart rate, or other bodily sensations \cite{glise_prevalence_2014}. \xl{According to embodied approaches, heightened awareness of bodily sensations can support individuals in identifying and making sense of their stress experiences.} Embodied methods, such as body scanning \cite{sauer2013comparing}, help individuals become aware of and identify these sensations or signals. By directly engaging people with their bodily experiences, these methods may serve as a powerful medium for recognising stress-related manifestations and using them as a foundation for communicating stress experiences and further interventions \cite{anne_cochrane_body_2022, yaribeygi2017impact}.} 
%\rh{The connection with adolescent girls lacks proper arguments. The following sentences should be further revised as there is no arguments there. We need to highlight WHY AD would benefit from embodied approach.} \xl{Adolescent girls, in particular, may benefit from such an embodied approach, as they are especially sensitive during adolescence in developing their body image. This heightened sensitivity means that body perception plays a more central role in their emotions, identity, and stress \cite{flaake2005girls}, making interventions that integrate bodily awareness particularly relevant to their developmental process.} \rh{After connecting embodied approach with AD, using one concise sentence to highlight the exact gap that motivates our current study.} 
\xl{While embodied approaches show promise for stress self-disclosure, little is known about how they can be leveraged to support adolescent girls' mental health, for whom bodily awareness is closely intertwined with emotional and identity development \cite{foster2025embodiment}.}

%Therefore, there is a potential that the body can serve as a powerful medium for exploring stress, facilitating expression, and navigating stress-related challenges \cite{yaribeygi2017impact}.}

%\hh{These} embodiment manifestations make the body a potential powerful medium for exploring and navigating stress \cite{yaribeygi2017impact}. Embodied approach, such as body scanning \cite{sauer2013comparing}, directly engage the individuals with their bodies, and offer unique opportunities to help them recognise the physical manifestations and use them as a foundation for self-disclosure, and further interventions \cite{anne_cochrane_body_2022}. 

\hh{In this study, we view the body as a a key medium for both understanding and expressing stress, and aim to investigate how an embodied approach can facilitate stress-related self-disclosure among adolescent girls. To explore this, we conducted two workshops to examine how to trigger the awareness of bodily sensations through embodied methods and how self-disclosure can emerge through the expression of bodily experiences. The first workshop, a pilot study, served as an initial exploration into whether and how bodily sensations could act as materials for self-disclosure. It also sought to understand adolescent girls’ expectations regarding the potential of such embodied approaches to facilitate sharing about stress. Insights from the pilot informed the design of the main workshop, Embodied Probes. It introduced a series of activities designed to help adolescent girls connect with their bodily sensations and support their expression through verbal, visual, and tangible forms. In addition, it encouraged participants to creatively explore ways of caring for their stressed bodies, further fostering self-reflection and self-disclosure}. Through this main workshop, we aimed to address the following research questions (RQs):

%To this end, we conducted two workshops with adolescent girls as participants. 
%\xl{Recognising that stress often manifests through bodily sensations—such as muscle tension, heartbeat, or breathing patterns—this study explores how such physical cues can be harnessed to support adolescent girls in articulating their stress experiences.} 

%how bodily sensations (e.g., stress symptoms) could support meaningful stress-related self-disclosure. Insights from this pilot informed the design of the main workshop, Embodied Probes, which expanded upon these initial findings through a structured series of activities focused on bodily experience. These activities included engaging with the stressed body, disclosing bodily sensations through verbal, visual, and tangible modalities, \xl{and creatively exploring ways to care for their own stressed bodies.} \rh{We need consider if ``designing interventions'' should be mentioned in current way. seems that it cannot inform our findings for the following RQs directly? Besides, the relationship between pilot and main, can be further clarified.} Through this main workshop, we aimed to address the following research questions (RQs):
\begin{itemize}
    \item \textbf{RQ1} How can embodied approach\hh{es} facilitate stress-related self-disclosure among adolescent girls?
    \item \textbf{RQ2} What dimensions of stress are disclosed by adolescent girls when using the body as a medium for self-disclosure?
\end{itemize}

Our work contributes to the HCI and DIS communities by 1) designing and iterating a series of embodied probes in workshops to facilitate stress-related self-disclosure, leveraging the body as an entity for feeling, communication, and design, 2) uncovering insights into the dimensions of stress that disclosed through embodied \xl{methods} among adolescent girls (i.e., somatic symptoms, sources, and coping strategies), and 3) offering design implications for future research on how an embodied approach can facilitate self-disclosure among adolescent girls and beyond. Based on our empirical findings, we argue for the potential of using the body as a medium to address mental well-being challenges, thereby contributing to the growing discourse on the “somatic turn” in third-wave HCI \cite{loke_somatic_2018}.

\section{Related Works}
\xl{In this section, we outline the motivation behind our approach. We begin by exploring the stressors faced by adolescent girls, highlighting the limited research on their mental health and the contextual impact of academic pressure in China. We then review existing self-disclosure practices in stress interventions, identifying a gap in the use of an embodied approach to support stress recognition and expression. Finally, we argue that embodied approaches, particularly soma design, offer a promising direction for addressing this gap.}
\subsection{\hh{Understanding Stress in Adolescent Girls}}
%\rh{This section should better address 2AC's concerns. consider re-organize it by: 1) starting from detailed introduction on ``common stressors or factors'' that affect AD's mental health or cause their stress, thereby highlighting our research is necessary and significant. 2) keeping the current second paragraph stating that there is little research for girls (highlighting Søndergaard et al.’s work on designing menstruation technologies with adolescents properly here). 3) highlighting that there is even less research in HCI for girls in China, and providing detailed backgrounds for readers to understand this population's stress. so that we can highlight the necessary of our current research.}

Stress and related negative mental states, such as anxiety and depression, can significantly disrupt adolescents’ daily lives, affecting their academic performance, social relationships, and overall mental well-being \cite{compas1993adolescent}. \xl{Common stressors that contribute to adolescents' stress include physiological (e.g. body development and hormonal changes)\cite{susman1988physiological}, psychological (e.g. low self-esteem and negative thinking patterns)\cite{fiorilli2019predicting}, social (e.g. peer pressure and parental expectations)\cite{mcmahon2020stressful} and cultural factors (e.g. cultural norms and stigma)\cite{lau2016adolescents}.} These challenges highlight the critical need for effective stress interventions. In response, the HCI community has increasingly focused on developing digital interventions to address adolescent mental health. A variety of technologies have been explored in this context, such as social robots \cite{rasouli_proposed_2022}, chatbots\cite{jusoh_helpbot_2024}, serious games \cite{catalan_storytelling_2024}, and virtual reality (VR) \cite{kitson_i_2024}. 

\xl{Although effective, existing research often treats adolescents as a homogeneous group, with little focus on the specific experiences of girls. This gap is clear in studies on emotional well-being and body-centred design. Søndergaard et al.’s work on menstruation technologies with adolescents is a key exception, showing how engaging with gender-specific, embodied topics can lead to more inclusive and supportive design\cite{sondergaard_designing_2021}.} In practice, significant gender differences exist in how stress is experienced during adolescence. Adolescent girls are more likely to report severe stress-induced symptoms, which, in adults, would be classified as chronic stress \cite{schraml_stress_2011}. For instance, eating disorders are four times more prevalent among adolescent girls compared to boys of the same age \cite{noauthor_mental_2017}. These differences arise from distinct sources of stress faced by girls, rooted in both physiological and social factors. On the physiological side, girls experience unique stressors related to hormonal changes during puberty, such as the onset of menstruation, which can lead to body image concerns and heightened emotional reactivity \cite{mccabe2003sociocultural}. Socially, adolescent girls often face expectations to conform to beauty standards or excel in caregiving roles, which can amplify feelings of inadequacy and stress \cite{gustafsson2009perceived}. 

\xl{Despite these gendered challenges, there is even less research in HCI that focuses on girls in non-Western contexts, particularly in China. This highlights the need for our current study, which seeks to better understand the stress experiences of Chinese adolescent girls through a culturally and socially situated lens. Academic stress is a significant and pervasive experience in the lives of Chinese adolescent girls, often shaped by high expectations, intense competition, and long study hours\cite{sun2012academic}. As a legacy of the one-child policy, many adolescent girls, particularly in urban areas, bear the full burden of parental expectations, with academic success frequently perceived as a reflection of family honour \cite{tsui2002only}. The highly competitive education system, centred around high-stakes exams such as National College Entrance Examination \textit{(Gaokao)}, places significant strain on adolescent students, resulting in long school hours and intensive after-school tutoring\cite{sun2013educational}.} 

\xl{It is therefore crucial to explore the specific characteristics of stress experienced by adolescent girls and to design targeted forms of support that are sensitive to their physiological, social, and cultural contexts. One promising pathway for such support lies in practices of self-disclosure, which can play a powerful role in stress intervention when approached through the lens of embodied design.}
\subsection{Self-disclosure for Stress Intervention}
There is a long-standing tradition in psychology research demonstrating that self-disclosure can help mitigate negative emotions and their potential adverse effects \cite{farber_self-disclosure_2006}. By allowing individuals to feel and express negative emotions, disclose relevant events or experiences, and reflect on their causes and potential coping strategies, self-disclosure serves as a therapeutic mechanism. 

In parallel, prior HCI research has extensively explored ways to facilitate such self-disclosure as a means of supporting adolescents in navigating mental health challenges. For example, Li et al. employed social robots to facilitate self-disclosure interactions with adolescents \cite{li_tell_2023}, while Pinch et al. focused on social platforms to support sensitive disclosure among sexual minority adolescents \cite{pinch_subtleties_2024}. Park et al. used conversational agents to reduce the burden of self-disclosure for victims of sexual violence \cite{park_can_2020}. \xl{Shi et al. developed an interactive sandbox designed to support left-behind children in rural China, enabling them to express themselves freely through playful and creative interaction.\cite{shi_disandbox_2024}. }

\xl{Despite the proven value of these research practices, verbal self-disclosure alone often neglects bodily sensations and falls short in fully capturing the depth of an individual’s emotional experience.} As mentioned earlier, mental well-being and the body are deeply interconnected; mental stress often manifests through physical symptoms \cite{glise_prevalence_2014}. For instance, stress can cause headaches, muscle tension, or gastrointestinal discomfort \cite{waldie2001childhood, edman2017perceived}. Few studies have explored embodied methods that encourage individuals to connect with their bodies, express bodily sensations, and communicate stress-related manifestations. 

For adolescent girls, stress often exacerbates bodily symptoms \cite{schraml2011stress}. These manifestations may include challenges related to menstruation, such as irregular cycles or heightened premenstrual syndrome (PMS) symptoms, underscoring the necessity of involving the body as a channel to explore and address stress. At the same time, menstrual-related stress is often difficult to disclose due to taboos and stigma surrounding menstruation \cite{campo_woytuk_touching_2020}. Andalibi et al. highlighted how young women frequently avoid disclosing stigmatised experiences, such as those related to reproductive health, on social media \cite{andalibi_disclosure_2020}. These barriers to open self-expression underscore the importance of creating safe and supportive environments where adolescent girls can explore and communicate their embodied experiences of stress.

Given the embodied nature of stress, such an approach holds promise for fostering self-disclosure, particularly among adolescent girls. Motivated by this gap, our study aimed to explore an embodied approach that leverages the body as a medium to probe and support stress-related self-disclosure in adolescent girls.

\subsection{\xl{Soma Design as an Embodied Approach}}
%\rh{I would suggest that shifting this section with 2.3. i.e., leaving it as the last section of our paper as typically the last section should be the one that is most relevant to our core contribution. Besides, I suggest that incorporating ``soma design'' in the this section's title to further address R1's concern. Regarding the detailed descriptions: 1) starting with embodied approach and its relationship with ``soma design'', and their relationship with ``body''. 2) then introducing how ``body'' was leveraged in previous study (you can organize the logic according to GPT's summary we saw last time). and highlighting that ``probe'' is important to help people builds connection/awareness with their body. 3) summarizing that soma design/embodied approach is promising yet under-explored for our contexts}

%HCI researchers have developed systems to facilitate the process of perceiving stress via embodied approach \cite{xue_co-constructing_2023}. Wearable technologies are used to monitor multiple modes of bio-feedbacks indicates stress, such as HRV(Heart Rate Variability) \cite{yu_biofeedback_2018} and GSR(Galvanic skin response) \cite{umair_thermopixels_2020}. Although embodied methods have been applied to address stress in various contexts, their potential to support stress self-disclosure remains largely unexplored. 

With the emergence of “somatic turn” in third-wave HCI, interaction design is incorporating embodied approaches such as soma design to explore bodily experience \cite{loke_somatic_2018}. \xl{An embodied approach in design emphasises the lived, felt experience of the body as central to understanding and shaping interactions with technology\cite{dourish_where_2001}. Soma design builds upon this perspective by engaging designers and participants in deep, reflective bodily engagement throughout the design process. Rather than relying solely on cognitive or verbal expression, soma design invites individuals to attune to subtle bodily sensations, using these as sources of insight and inspiration\cite{hook_designing_2018}. Both perspectives recognise the body as a site of knowledge and expression, positioning it as an essential participant in the design process, particularly when addressing intimate or sensitive topics such as stress or well-being.}

%\xl{High levels of stress can lead to somatic symptoms, where psychological distress is expressed through physical discomfort or bodily sensations\cite{glise_prevalence_2014}. For adolescents, who may struggle to articulate their emotions verbally, these bodily expressions often become an important mode of signalling distress\cite{bohman2010somatic}. In this context,} designing for tangible interaction \cite{hornecker_getting_2006} has been shown to be effective in supporting adolescent to communication about their mental well-being when facing emotional challenges. A recent project highlights the potential of providing real-time contextual support for emotional challenges and integrating embodied cognition in designing \xl{tangible} interventions for adolescent mental well-being \cite{dauden_roquet_exploring_2022}. Additionally, wearable devices have been designed to capture users' stress levels, which can be shared with peers online \cite{jiang_intimasea_2023-1}, showing the potential of \xl{embodied} technologies for facilitating stress self-disclosure.

 \xl{Soma design is characterised by its focus on bodily sensation as the starting point, guided by reflective bodily explorations, and grounded in enactive practices that deepen bodily awareness\cite{hook_designing_2018}. We draw inspirations from soma design research projects led by Søndergaard et al., who facilitated embodied ideation in the context of interaction design for adolescent girls' sensitive menstrual experiences through an participatory and embodied approach\cite{sondergaard_designing_2021}, and from Segura et al., who introduced embodied sketching, a practice that engages participants’ situated bodily experiences to generate and explore design ideas\cite{marquez2016embodied}. These prior works informed our understanding of how an embodied approach can support self-disclosure, particularly in sensitive contexts, and laid the foundation for our application of soma design in addressing stress.}

\xl{Soma design often relies on the use of embodied probes to access and explore individuals’ lived bodily experiences. Unlike traditional cultural probes \cite{gaver_design_1999}, design probes \cite{wallace_making_2013} and material probes \cite{jung_material_2010}, embodied probes are specifically designed to elicit bodily sensations, helping participants reflect on how experiences are felt through the body. Commonly used probes in soma design include} body-based practices like body mapping \cite{anne_cochrane_body_2022} and tangible body maps \cite{nunez-pacheco_dialoguing_2022}, enabling closer self-observation and awareness of the body, demonstrating the intertwined nature of embodied interaction  \cite{hook_unpacking_2021}. \xl{Soma design guided the development of our workshop activities, acts as a bridge to self-disclosure by first helping participants become aware of their bodily sensations, then encouraging them to express these experiences through various forms, ultimately connecting embodied feelings with personally meaningful solutions.}

Although there is increasing interest in HCI regarding the communication of embodied experiences via creative and tangible methods \cite{claisse_tangible_2022}, more research on how to leverage the body as a proactive agent to promote mental well-being is still needed. \xl{The use of soma design as an embodied approach for stress self-disclosure,} while promising in HCI contexts, has been insufficiently explored. This study aims to contribute to existing research by exploring a soma-based approach to promoting stress-related self-disclosure through embodied methods with adolescent girls.

\section{Overall Methods}
\subsection{Research Aim and Methods Overview}
\noindent
\xl{The} overall aim of our research is to support adolescent girls in stress self-disclosure through an embodied approach, which leverages the body as a medium \hh{for understanding and expressing stress}. To achieve this goal, we first conducted a pilot workshop with six adolescent girls (aged 18–19) to explore potential embodied methods that could inform this approach (see \autoref{pilot}). Through this workshop, we gained key insights into: (1) designing more detailed body-examining experience to support participants to fully engage with their bodies; (2) incorporating multiple modalities to better express bodily sensations; and (3) introducing diverse activities and tools to help them craft interactive solutions for their stressed bodies.

Building on these insights, we iterated and implemented Embodied Probes (see \autoref{main}), a co-design workshop involving nine adolescent girls (aged 15–18). Participants engaged with their stressed bodies through body-scanning instructions, expressed their bodily sensations using body mapping (in visual forms), tangible materials, and verbal communication, and leveraged their embodied experiences to address personal stress using an interactive toolkit. Throughout this process, we collected various materials created by participants, including drawings, clay sculptures, wearable prototypes, and interview data. 

Finally, we analysed the collected materials using thematic analysis to address our research questions (see \autoref{findings}). For RQ1, our findings demonstrated how the workshop facilitated a series of body-centric experiences that supported self-disclosure. These experiences included engaging with and feeling bodily sensations as a foundation for self-disclosure, enabling self-expression through multimodal channels, and leveraging the crafting process to delve deeper into why and how adolescent girls address stress using their bodies as a medium. For RQ2, we uncovered the somatic symptoms, sources, and coping strategies of stress that were disclosed during the workshop. These findings serve as the foundation for the design implications discussed in \autoref{discussion}.

\subsection{Ethical Considerations for Participants Recruitment}
For participant recruitment, we first adhered to the WHO’s definition of adolescence as the life phase between 10 and 19 years old. Given that the likelihood of mental health disorders peaks in the 15–19 age group \cite{noauthor_mental_nodate}, we focused our study on adolescent girls within this age range. \hh{As mentioned earlier}, in China, girls in this group often face significant academic pressure, especially as they prepare for the college entrance exam, typically taken at the end of their third year in high school. % Most of the participants in our study were between 15 and 18 years old, aligning with this critical developmental and academic period.

Since our target group included participants under 18 years old and the research involved sensitive topics, we placed particular emphasis on ethical considerations throughout the recruitment process. We developed a research protocol in close collaboration with the partner high schools, ensuring that parental consent for participants under 18 was obtained through the schools. We prioritised participants' consent and agency. Before each workshop, we thoroughly explained the study protocol to all participants and obtained their informed consent, \hh{and also confirmed that none of the participants had identified mental disorders. After each workshop, we maintained close communication with the schoolteachers to ensure the participants' wellbeing.} This study received ethics approval from the first author’s institution.

%\rh{we should NEVER say so like ``did not screen...disorder'' in our paper, as it may cause sever ethical concerns. According to our IRB, our participants should not have any identified mental disorders. So I suggest that we should remove the following descriptions (I had incoperated the necessary info into the above texts).}
%\xl{We did not screen participants for potential body image disorders prior to the workshop. However, we maintained close communication with the schoolteachers throughout the process and conducted follow-up interviews afterwards to ensure the participants' wellbeing and to address any concerns that may have arisen.}

% First of all, in China, most students are required to take the college entrance exam at the end of their third year in high school, which creates a unique peak in academic stress around the age of 18. The academic pressure also limited the time adolescents could dedicate to participating in this study. Considering all these factors, we decided to focus the main study on adolescent girls aged 15-18 years in high school. 

\subsection{Positionality}
The authors of this research are cisgender women, aged between 21 and 36, from China and the UK. All authors have personal experience as adolescent girls and have encountered various forms of stress. For most authors, their formative years took place in China, allowing them to share a similar cultural background with many participants. This shared context, along with past experiences of stress, helped us relate to and empathise with the participants in our workshops. To further enhance our understanding, we engaged in meditative exercises before and during the research process to increase our bodily awareness and better comprehend the body-related self-disclosures shared by the participants. We also reflected on how our past experiences, particularly the intense academic stress we once faced, might unintentionally lead to an overemphasis on academic stress in our interpretations. To mitigate this potential bias, we held regular meetings to discuss and critically reflect on how our positionality might shape our findings, ensuring a more balanced and nuanced analysis.

% Throughout the research process, we reflected on our own stressful experiences during adolescence. Authors from China, in particular, focused more on academic stress, as adolescents in China often face higher academic expectations due to the cultural context. Our current and past experiences of stress helped us relate to and empathize with the participants in our workshops. 

% All authors also share a similar background to the participants, who aspire to become designers themselves.  

\section{Pilot Workshop}
\label{pilot}
\xl{This section presents a pilot workshop with female adolescents that served as a preliminary study. Insights from this workshop informed the development of the Embodied Probes workshop, which aimed to facilitate peer-to-peer sharing and sense-making around embodied experiences of stress.}

% we briefly describe our initial research activities: a pilot workshop with female adolescents, which served as pilot study. Insights from pilot workshop informed our approach for the Embodied Probes workshop – a method we developed to facilitate peer-to-peer sharing and sense making of embodied experiences of stress.
\subsection{Workshop Protocol}

\subsubsection{Participants}
Six adolescent girls (aged 18–19, all self-identified as cisgender women) participated (named M1 to M6) in our pilot workshop. The participants were recruited through an open call posted on a popular social media platform in China. Each participant received a compensation of 30 Chinese Yuan upon completing the workshop.
% The pilot workshop took place during students’ final week of the semester, which likely influenced participants’ levels of stress. 

\subsubsection{Procedures}
The workshop took place in a lab space featuring a large table where workshop materials were arranged. Participants sat in a circle around the table, fostering a sense of community and open communication. Two authors co-facilitated the workshop, introducing the procedures and protocols to all participants at the beginning. The workshop lasted 90 minutes and consisted of the following three sessions (see \autoref{fig2}):

% The workshop was held in a lab space, featuring a large table where workshop materials were placed. The workshop protocol was as follows: each participant was invited to sit in a circle and fill in a questionnaire about stress, which serves as a brief introduction to the workshop aim. 

\xl{We followed a typical soma design pipeline comprising discovery, appreciation, and design, guiding participants to first notice their bodily sensations, then deepen their awareness through reflection, and ultimately translate these insights into tangible design outcomes.}

\textbf{\textit{Session 1. Identify Stress Symptoms:}} Participants were encouraged to use a wooden artist figure and pink dotted stickers to mark the areas where they experienced stress symptoms. As they placed the stickers, they shared their stress symptoms with the group. The purpose of this session was to help participants explore their bodily sensations of stress and encourage their self-disclosure simultaneously.

\textbf{\textit{Session 2. Share a Stress Event Relating to Symptoms:}} The participants were then asked to share a recent experience of stress relating to the symptoms identified in Session 1 in detail, including the sources of stress and their coping strategies. After sharing, they were prompted to ring a crystal singing bowl to help restore mental wellness. This session was designed to explore whether the exploration of bodily sensations could better support self-disclosure of stress. 

\textbf{\textit{Session 3. Curate Stress-relief Method:}} Finally, participants had the option to curate and engage in a stress-relieving activity using the probes, which included materials such as collage supplies, sheets of felt, clay, sewing tools, and more. This activity encouraged participants to design personalised stress-relief methods for the stressed bodies they had identified.

\subsubsection{Data Collection and Analysis}
\begin{figure*}[h]
    \centering
    \includegraphics[width=1\linewidth]{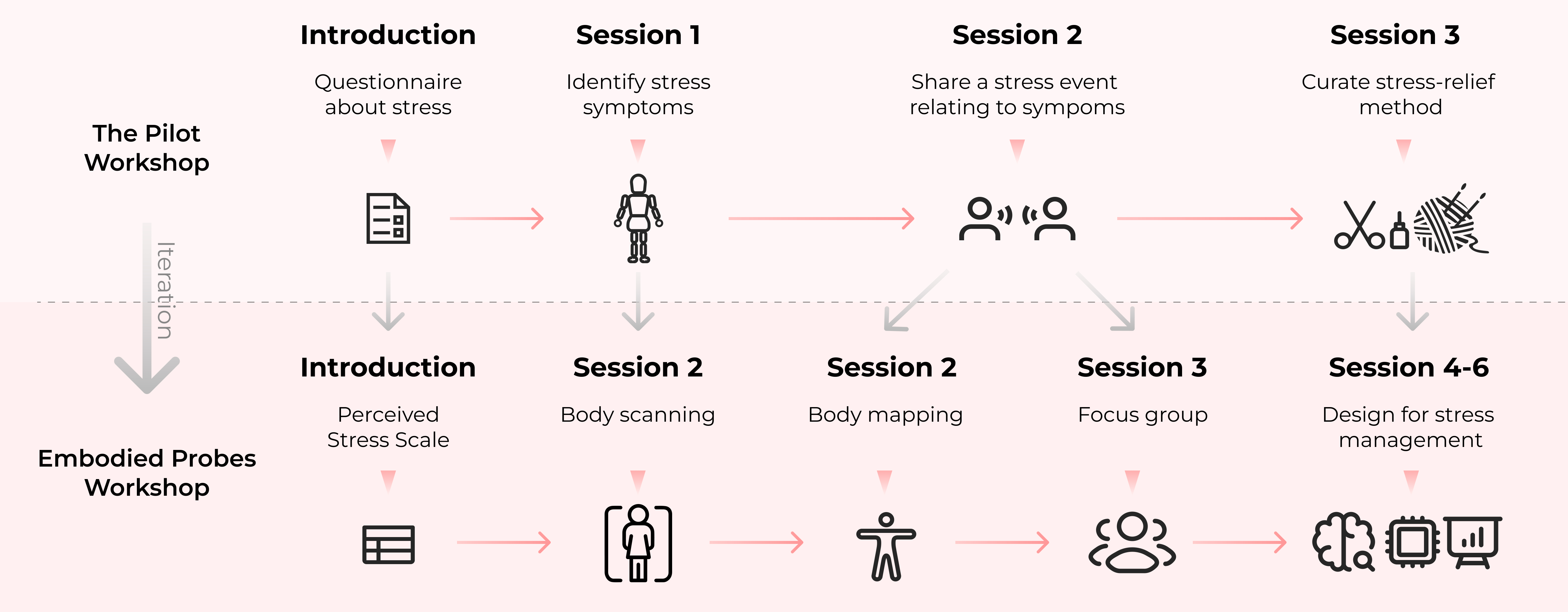}
    \caption{Iteration between pilot workshop and Embodied Probes workshop.}
    \Description{Fig. 1: A flowchart comparing the structure of the Pilot Workshop and the Embodied Probes Workshop. The Pilot Workshop includes an introduction (questionnaire about stress), Session 1 (identify stress symptoms), Session 2 (share a stress event), and Session 3 (curate stress-relief method). Below it, the Embodied Probes Workshop shows an iteration based on the pilot, with an introduction (Perceived Stress Scale), Session 2 (body scanning), Session 2 continued (body mapping), Session 3 (focus group), and Sessions 4–6 (design for stress management). Icons visually represent each session's theme.}
    \label{fig2}
\end{figure*}

The workshop was video-recorded to capture participants' thoughts on the workshop activities and the stress-related information they self-disclosed. Following the workshop, individual interviews were conducted and audio-recorded to gather participants' reflections on the workshop and their perceptions of their bodies during and after the session. All recordings were transcribed, and an open coding process was independently conducted by two authors. Through an iterative process, discussions were held to reach consensus. The analysis focused on participants' comments and suggestions for each session to better inform the iteration and design of the main workshop (\autoref{fig2}). 

\subsection{Findings from Pilot Workshop}
First, participants' feedback indicated that using the body as a medium to explore and disclose their stress was an interesting and novel experience that they had not encountered before. They also expressed enthusiasm and expectations for future workshops of this kind. Additionally, our analysis identified three overarching themes related to the three workshop sessions, which were crucial in shaping the design of our subsequent workshop. 

\subsubsection{Expectation for a More Detailed Experience to Feel the Body} In Session 1, we used a wooden artist figure to help participants examine different areas and relate them to their own bodily sensations. This approach allowed participants to engage with and feel their bodies, helping them identify somatic symptoms of stress they had not noticed before. For example, M2 noted: \textit{“I've never really paid attention to my body like this before. I noticed discomfort in several areas, especially my neck. I used to think it was just because of poor sleep.”} However, participants also suggested incorporating a more detailed examination of their bodies. For instance, M4 shared: \textit{“Focusing on my body helped me better understand and share my feelings. Connecting with the physical sensations allowed me to reflect on what might have caused the discomfort. Listening to others talk about their stress also helped me gain a clearer understanding of my own bodily experiences.”}

\subsubsection{\hh{Expectations for More Channels to Support the Expression of Bodily Experience of Stress}} We found that during Session 2, participants tended to share stress events that were related to the bodily sensations they identified in Session 1. For example, M1 shared: \textit{“When I have a lot of assignments, I stay up late. The lack of sleep causes me to experience chest pain and palpitations.”} However, many participants expressed that it was difficult and challenging to fully articulate how their bodies felt and to clearly convey their inner feelings and thoughts \hh{verbally}. For instance, M4 said: \textit{“I don't know how to say this, but there's a strange feeling in my wrist. I need to sit here and twist it a bit to feel more comfortable.”} Similarly, M3 mentioned: \textit{“I know that every time I’m stressed out, I get back pain. But it’s hard to describe the feeling through words, and sometimes it’s difficult for me to locate the exact area of the pain. Often, my entire back hurts.”} These comments inspired us to consider incorporating more nuanced methods to help participants better articulate and express their bodily experiences of stress.

\subsubsection{Expectations for Tangible Interactions with Bodies through Self-Crafting} In Session 3, we found that craft-based activities were effective in relieving stress for our participants. Specifically, activities such as sewing, sculpture-making, and needle felting were reported as useful methods. For example, M3 noted: \textit{“During the collage-making process, I felt very immersed and engaged, which helped relieve my stress.”} In addition, participants expressed their expectations for interactive prototypes that could engage more directly with their bodies. M5 shared: \textit{“I hope there could be smarter wearable devices to detect my stress. The smart band I use now doesn't work well for me; whether I'm relaxed or stressed, it always shows a moderate level.”}%“I wish I could knit a colour-changing or shape-changing blanket that would follow my breathing so I could feel my belly rise and fall. By observing the rise and fall of my stomach, I sometimes feel that my stomach does not hurt as much. Such activities would make me feel relieved.”

In summary, the pilot workshop helped us identify key design considerations for the following Embodied Probes workshop: (1) designing more immersive and detailed body-examining sessions to encourage participants to fully engage with their bodies; (2) incorporating additional channels to support the elicitation of bodily sensations and related stress experiences; and (3) integrating more interactive and tangible tools to enable participants to craft personalised stress-relief solutions.

\section{EMBODIED PROBES WORKSHOP}
\label{main}
\hh{Based on the findings from our pilot workshop, we further developed the main workshop named Embodied Probes. By facilitating more detailed examinations of bodily sensations and offering more channels for expressing these sensations, the main workshop aims to: (1) scaffold how embodied approaches can support adolescent girls in stress-related self-disclosure, and (2) reveal richer empirical insights into the stress experiences of this group.}

%\xl{The goal of the Embodied Probes workshop was to facilitate a somatic, embodied experience, enabling participants to explore how stress manifests in their bodies, self-disclose their stress and to design responses that address it.}
\subsection{Participants}
Information about the workshop was published on a social media account run by a local high school. The call was open to any adolescent based in Shanghai who identified as female and was aged 15-18 years old. A snowball sampling method was initiated by the Deputy of Art from the school, and a total of nine participants were recruited \hh{(P1-P9)}. \xl{The participants were from three different school years and are partially familiar with each other depending if they were in the same year. They attend school from 8 a.m. to 5 p.m. on weekdays and typically have at least four hours of homework after school.} They also take extra lessons during weekends and holidays. At the end of the workshop, participants were given the opportunity to choose a gift they liked, valued at 50 Chinese Yuan.

\subsection{Workshop Design and Protocols}
%Design considerations from the pilot workshop informed the iteration of our design for the 
\subsubsection{Overall Considerations}
\hh{The main workshop was conducted over two days with nine participants and was facilitated by the first author along with four co-authors. In designing the workshop, we carefully considered the characteristics of adolescent girls and the goals of our study. We emphasised that the purpose was to listen and understand, not to evaluate, and participants were given full autonomy over what they chose to share at any time. To create a safer and more comfortable environment for open expression, no boys were included in the workshop, reducing the influence of the male gaze and the potential discomfort around discussing embodied or taboo experiences. Additionally, two co-authors close in age to the participants were involved to foster relatability and reduce the perceived authority gap. To avoid associations with academic stress, the workshop was held in a lecture theatre that was intentionally arranged to feel more relaxed and less like a traditional classroom.}

%We intentionally involved two of authors, who were of a similar age to the participants, which facilitated easier bonding with them. \xl{In designing the workshop, priority was given to avoiding factors that might trigger body image concerns—such as the influence of the male gaze—by ensuring no male participants were involved. To ensure participants felt well cared for and comfortable, breakfasts, light snacks and lunches were provided throughout the workshop. We were mindful that food-related situations can be sensitive, especially for girls who may have concerns around eating or body image. We provided a variety of food options, including familiar, non-triggering, and non-restrictive choices, without drawing attention to eating behaviours. Considering that a traditional classroom environment might trigger chronic academic stress for some participants, the activities were held in a lecture theatre, which was intentionally arranged to create a more relaxed and less classroom-like atmosphere. For instance, participants were encouraged to sit in a circle on the floor, fostering a sense of openness and equality, and yoga mats were provided to support body scanning exercises and other relaxation activities.}
\begin{figure*}[h]
    \centering
    \includegraphics[width=1\linewidth]{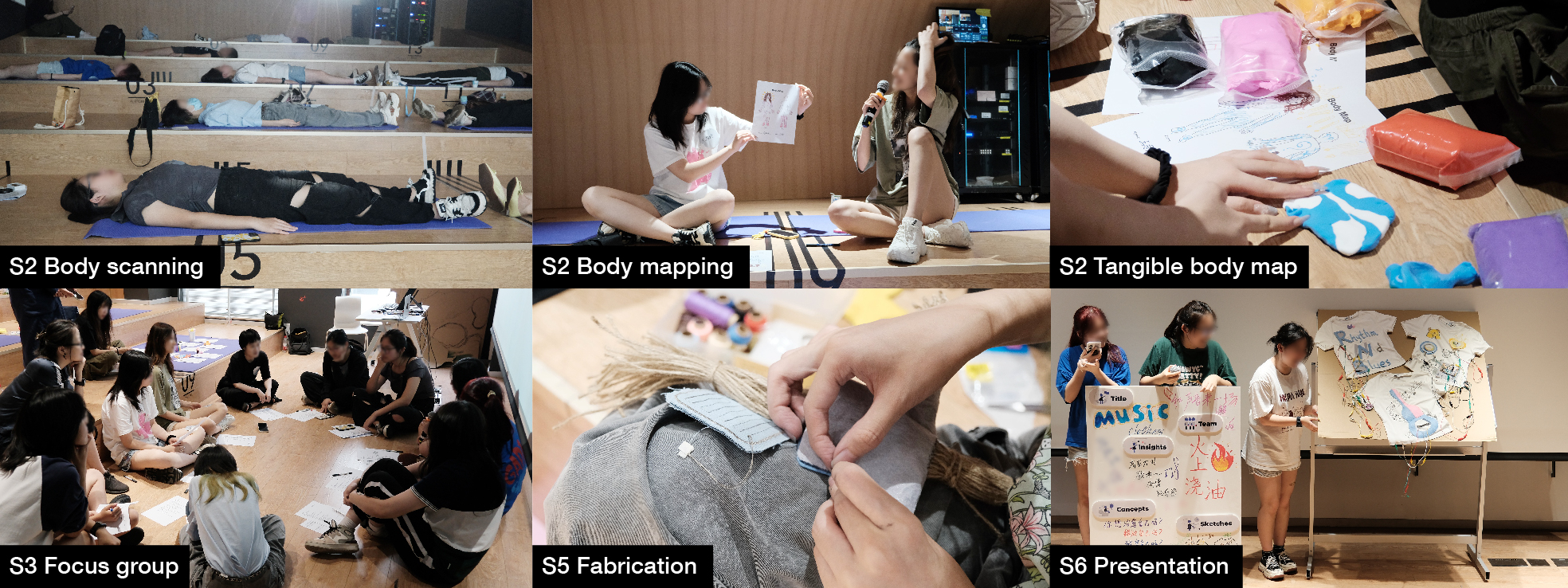}
    \caption{Embodied Probes Workshop Sessions.}
    \Description{Fig. 2: A collage of six photos depicting various activities from the embodied probes workshop. Top left shows participants lying on yoga mats for body scanning. Top middle shows two girls presenting during a body mapping session. Top right shows a close-up of a tangible body map with clay pieces. Bottom left shows participants sitting in a circle during a focus group discussion. Bottom middle shows hands sewing fabric during a fabrication session. Bottom right shows three girls presenting their final design of three customised t-shirts alongside a poster.}
    \label{fig3}
\end{figure*}

\subsubsection{Workshop Protocols}
\hh{Guided by the overall considerations mentioned above, we structured the workshop into a series of detailed sessions, accompanied by specific design considerations.}

\hh{\textbf{\textit{Session 0. Warm-up:}} Before the official start of the workshop, the facilitators arrived early to prepare the space and welcome participants as they arrived. During this informal period, facilitators engaged in casual conversations with participants, such as introducing ourselves, asking about recent movies or books they enjoyed, and other everyday topics. As participants settled in, they were free to enjoy the snacks we prepared, and the conversations often happened while eating together. This gentle, unstructured interaction helped ease any initial nervousness and allowed facilitators to begin building a foundation of trust and familiarity. We intended to use this informal rapport to set a supportive tone for the rest of the workshop and contributed to a more relaxed and open atmosphere.}

\textbf{\textit{Session 1. Introduction:}}
\hh{Our workshop began by establishing a shared understanding. We first introduced basic concepts related to mental health and stress, emphasising that stress is a normal and valid emotional experience. We also introduced the idea of the body as an active participant in emotional expression, discussing how emotions are often manifested through bodily sensations. Throughout the introduction, we emphasised that all feelings and ideas were welcome, and that the workshop was a non-judgmental space.} Perceived Stress Scale (PSS-10) \cite{cohen_perceived_1988} was used to assess participants’ perceived stress levels prior to engaging with the Embodied Probes. 

%We presented the participants with a design brief to create a technology to perceive, communicate, or manage their stress. The goal was to giving participants a rough idea of the design space. Perceived Stress Scale (PSS-10) \cite{cohen_perceived_1988}was used to assess participants’ stress perceptions before the use of Embodied Probes. 

\textbf{\textit{Session 2. Body Scanning and Body Mapping:}} In this session, we aimed to help participants locate the sources of their stress and establish \hh{communication channels} with their bodies, therefore reconnecting with their stressed bodies and opening up themselves of self-disclosure. \xl{Inspired by the embodied probes used in soma design,} \hh{we first facilitated a body scanning exercise to support participants in connecting with and sensing their bodies. Recognising that adolescent girls may find it challenging to verbally articulate internal sensations or emotions, we incorporated body mapping \cite{anne_cochrane_body_2022} as a way for them to visually express their bodily experiences. Participants were also invited to create clay sculptures, a form of tangible body map \cite{nunez-pacheco_dialoguing_2022} based on specific felt experiences, allowing for a more tangible and material exploration of stress. These visual and tactile activities supported the processes of visualising,  reflecting, and expressing on internal feelings. We further detailed these embodied methods in \autoref{embodied_methods}.} 

%we first facilitated a body scanning exercise and used body maps as a generative tool \cite{anne_cochrane_body_2022} to support participants in communicating their patterns of bodily experiences of stress. To encourage participants to further reflect on their bodily experiences of stress, we drew from the method of tangible body maps and asked them to create a clay sculpture based on a specific bodily experience \cite{nunez-pacheco_dialoguing_2022}. Upon finishing the drawings and sculptures, the participants share their work and the surfaced bodily experience to the group. 

\textbf{\textit{Session 3. Focus Group:}} \hh{After helping participants tune into their bodies through a body-scanning exercise and encouraging them to reflect on their felt bodily sensations, they were guided into a focus group discussion to share their thoughts and feelings around stress. To create an open and safe communication environment, we emphasised that there were no right or wrong answers, and that all experiences were valid. Participants were reminded that they could choose how much to share, and that listening was just as valuable as speaking.} To create an open and safe communication environment, the nine participants were invited to sit in a circle during this session. We followed practices for conducting research with adolescents and encouraged our participants to take on the role of co-researchers \cite{lygnegaard2023adolescent}. The focus group was divided into three parts and in each part, participants switched between the roles of facilitator, interviewee and observer. The goal here was to encourage the girls to practice qualitative inquiry (e.g., focus groups) with their peers, thereby developing their confidence and encourage them to conduct their research on their own one day. \xl{The focus group discussions centred around the girls’ personal experiences of stress, how stress manifests in their bodies, the contexts in which these sensations arise (such as school or family settings), and the ways they currently cope with or respond to these embodied feelings.}

%With the intention of empowering the adolescent girls with not only with tools but also methods for self-disclosure, we designed a safe and intimate setting for facilitating our focus group. 

\textbf{\textit{Session 4. Brainstorming:}} \xl{Drawing upon the embodied ideation process engages young participants\cite{marquez2016embodied}, the focus was on using design activities as a means to facilitate self-disclosure for stress, rather than as a goal in itself. The session aimed to help participants engage with and express their experiences of stress in a tangible way.}  Each group was given a sheet of A0 paper to draw their concept maps and sketch their ideas. A \xl{series of games were played before} the standard routine of brainstorming in design thinking was followed \cite{noauthor_waht_nodate}. By the end of the session, participants selected their top three ideas, which were further developed on the second day of the workshop. 

\textbf{\textit{Session 5-6. Design for Stress Management (Fabrication and Presentation):}} \xl{Inspired by a participatory soma design project in which adolescent girls designed menstrual technologies for their newly-menstruating bodies \cite{sondergaard_designing_2021}, the goal of the final sessions was to offer adolescent girls an opportunity to realise their ideas for addressing the stress they perceived. The design activities were used as a means to deepen self-disclosure and explore coping strategies.} By the end of day two, the three groups had presented fully functioning prototypes, accompanied by their posters. The presentation concluded with each participant awarded with a certification of design and research skills that they learned at the workshop.

\textbf{\textit{Follow-up Interviews:}} Follow-up interviews of the Embodied Probes workshop were conducted by three authors with all workshop participants in a café near the school campus a week after the workshop. \xl{The aim was to evaluate both the overall workshop design and the embodied methods used throughout the two-day workshop.} The initial set of questions teased out general feedback about the two-day workshop, including inquiries about their favourite sessions or activities and the reasons behind their choices, as well as their reflection on the workshop experience. 

\subsection{Embodied Probes in the Workshop}
\label{embodied_methods}

\xl{Embodied probes are tools and practices designed to elicit bodily sensations and foster reflective awareness, serving as a means to access, explore, and express participants’ lived experiences of stress.}
\subsubsection{Body Scanning} \hh{Participants started to connect with their bodies through body scanning. We emphasised that there were no right or wrong feelings, encouraging participants not to judge themselves during the scanning process. Then, the participants were invited} to lie down on yoga mats. They were then guided through a body scanning exercise via a pre-recorded audio\footnote{https://b23.tv/qqE9sbn}. The soft lighting and calm tone of the audio were designed to foster a sense of relaxation, allowing participants to let their guards down and become fully at ease. \xl{The audio lasted for 13.5 mintues, and it involved the whole body scanning.} %Throughout the audio,  %As it was the first interactive session of the workshop, it also served as an ice-breaker to help participants feel more comfortable and engaged. 

% First of all, author 1 provided a brief  overview of the concept of body mapping and the process for creating a body map \rh{<- can we leave the mapping-related descriptions to next subsection? otherwise, it will confuse readers} before inviting participants 

\subsubsection{Body Mapping and Tangible Body Mapping}
Based on the detailed examination participants engaged in during the body scanning, they were then encouraged to visualise their bodily experiences through drawing and share their creations with the group. Each participant was provided with a sheet of paper featuring front and back outlines of a human body, along with coloured pens. The body mapping exercise served three key objectives as a generative tool \cite{anne_cochrane_body_2022}. \xl{First, it allowed participants to focus on their bodily experiences, heightening body awareness. Second, it helped to establish a communication channel between participants and their stressed bodies. Finally, the bodily experiences captured through the body maps served as a source of inspiration for the participants in design and making process.}

Immediately after the creation and presentation of the body maps, we encouraged participants to focus on a specific bodily experience illustrated on their body maps and transform it into a sculpture. For this, we used an open-ended tool called tangible body mapping \cite{nunez-pacheco_dialoguing_2022}. Participants were provided with coloured clay and basic sculpting tools. Once they completed their sculptures, they shared their insights and reflections with the group, and follow-up questions were asked to explore their decision-making processes—such as their choices of colours, forms, and the content of the sculptures. Tangible body mapping enabled participants to delve deeper into their body awareness and evoke more meaningful content. This approach sparked a somatic dialogue, allowing participants to reinforce and express their nuanced bodily experiences through the act of crafting.
\subsubsection{Expressive Making} \xl{Following a brainstorming and ideation session inspired by prior embodied methods, participants were invited to give physical form to their personal reflections on stress. The goal of this hands-on activity was to use the act of making as a means of self-disclosure by externalising participants' internal experiences through design. To support this expressive process, we provided a range of materials, such as felt sheets, T-shirts, threads, sewing kits, and more. Participants were also encouraged to explore embodied technologies, such as wearables and tangible interactions, as expressive elements in their designs. Informed by Design Consideration 3 from the pilot workshop, the construction toolkit “EmTex” \cite{wang_emtex_2023}, featuring embroidered textile-based wearable modules, was made available to help prototype concepts. These smart textile sensors, along with technical guidance, enabled participants to explore how stress might be represented or sensed through the body. Overall, the focus of this session was on supporting participants in articulating their lived experiences of stress through creative, bodily-engaged forms of making.}

\subsection{Data Collection and Analysis}
\hh{Audio recordings from all Embodied Probes workshop sessions and follow-up interviews were transcribed and anonymised for analysis. In addition to verbal data, we collected visual and tangible artefacts created by participants, such as body map drawings, clay sculptures, posters, and concept maps. These artefacts were treated as part of the qualitative dataset and analysed alongside participants’ verbal descriptions and reflections.}

\hh{We employed an inductive thematic analysis approach \cite{braun_using_2006} to examine the full dataset. Two authors independently familiarised themselves with the transcripts and contextual notes accompanying the visual and tangible artefacts, then conducted initial coding separately. Throughout the process, the coders met regularly to compare interpretations, resolve discrepancies, and collaboratively refine a shared codebook. Through iterative discussions, related codes were grouped into broader thematic categories that captured key patterns in the data. Final themes and subthemes along with concise definitions are presented in the codebook (see Table~\ref{table1}). These include themes such as \textit{“safe space for bodily awareness”, “somatic symptoms of stress”} and \textit{“sources of stress}, etc.} Additional meetings were organised with Author 2 to review the themes in relation to participants’ accounts.

\section{Findings}
\label{findings}
\hh{In this section, we present findings and observations from the main workshop in response to our two research questions. For RQ1, we detail how an embodied approach can facilitate stress-related self-disclosure among adolescent girls. For RQ2, we identify the somatic symptoms, perceived sources, and coping strategies of stress that were disclosed by the participants.}

\begin{table*}
    \centering
    \begin{tabularx}{\textwidth}{p{4cm} X p{5cm}} % 第一列宽度改为4cm，确保能换行
        \toprule
        \textbf{RQs} & \textbf{Themes and Definitions} & \textbf{Subthemes} \\
        \midrule
        \multirow{4}{=}{\parbox[t]{4cm}{\textbf{RQ1:} Supporting Stress-related Self-disclosure through an Embodied Approach}} 
            & \textbf{Safe Space for Bodily Awareness:} An environment where individuals feel secure to explore and connect with their bodily sensations. & Sense of safety\newline Immersed experience \\
            \cmidrule(lr){2-3}
            & \textbf{Feel and Connect with Stress:} Using bodily awareness techniques, like body scanning, to recognise and engage with the bodily sensations of stress. & Rarely engage with bodily sensations\newline Aware of bodily experience \\
            \cmidrule(lr){2-3}
            & \textbf{Self-Disclosure of Stress:} Expressing and revealing stress-related bodily sensations using visual representations and tangible objects. & Visual or tangible forms\newline Metaphorical imagery  \\
            \cmidrule(lr){2-3}
            & \textbf{Body-Based Stress Intervention:} Using body-based interventions to address and reduce stress that has been identified or expressed. & Body as medium \\
        \midrule
        \multirow{3}{=}{\parbox[t]{4cm}{\textbf{RQ2:} Self-disclosed Dimensions of Stress}}
            & \textbf{Somatic Symptoms of Stress:} The physical manifestation or bodily expression of stress. & Chronic pain\newline Gastrointestinal symptom\newline Musculoskeletal symptom \\
            \cmidrule(lr){2-3}
            & \textbf{Sources of Stress:} Factors, situations, or events that trigger stress in individuals. & Academic stress\newline Interpersonal stress \newline Stereotype threat\newline Body image\newline Family dynamic \\
            \cmidrule(lr){2-3}
            & \textbf{Means of Relieving Stress:} Coping mechanisms or strategies individuals use to manage stress. & Relaxation object\newline Sensory modality\newline Creative activity \\
        \bottomrule
    \end{tabularx}
    \caption{The codebook}
    \Description{Table 1: A table presenting themes, definitions, and subthemes corresponding to two research questions (RQ1 and RQ2). For RQ1, themes include “Safe Space for Bodily Awareness,” “Feel and Connect with Stress,” “Self-Disclosure of Stress,” and “Body-Based Stress Intervention.” For RQ2, themes include “Somatic Symptoms of Stress,” “Sources of Stress,” and “Means of Relieving Stress.” Each theme includes a definition and a list of subthemes, such as “Sense of safety,” “Visual or tangible forms,” “Chronic pain,” and “Relaxation object.”}
    \label{table1}
\end{table*}

\subsection{\hh{Supporting Stress-related Self-disclosure through an Embodied Approach (RQ1)}}
\hh{In our workshop, the embodied approach was implemented through various methods such as body scanning and visual or tangible body mapping. Our analysis revealed how these methods collectively supported stress-related self-disclosure among adolescent girls. Specifically, the process of self-disclosure began with the creation of a safe space where participants could explore and connect with their bodily sensations. After helping them engage with their bodily sensations through scanning, the body mapping activities prompted them to reflect on their embodied experiences, which in turn facilitated verbal expression of their stress-related thoughts and feelings. Finally, by encouraging participants to care for their bodily experiences through hands-on crafting activities, they were further supported in articulating their desired ways of coping with stress. Below, we present our detailed findings.}

%The embodied approach to self-disclose stress begins by creating a safe space where participants can explore and connect with their bodily sensations. Once aware of their stress, they can express it through visual or tangible means. This self-disclosure allows for a deeper understanding of the stress, which is then addressed through body-based interventions to relieve stress.
\subsubsection{\hh{Encouraging Bodily Awareness Through Safe and Immersive Experiences}}
\hh{Working with the body can sometimes make people feel vulnerable. To foster \textbf{a sense of safety} in the workshop, we deliberately included only girls as participants. Our analysis revealed that such consideration created a comfortable environment where participants would feel more at ease. For example, P8 shared in the follow-up interview, ``\textit{If there are boys around, I would actually feel more stressed, but with just girls, I would feel much more at ease.}'' Feeling safe was foundational as it help participants begin to open up and reconnect with their bodies.}
%Working with the body sometimes makes people feel vulnerable, therefore in the workshop settings we first considered to have only girls as participants to support the \textbf{sense of safety}. \textit{“If there are boys around, I would actually feel more stressed, but with just girls, I would feel much more at ease.”} (P8, follow-up interview) Only when the participants feel safe can they open up and connect with their bodies, which is the foundation of the workshop.

\hh{Besides, we found that the dimmed lights, yoga mat, quiet environment that we carefully curated during the body scanning, created a calm atmosphere that encouraged} fully \textbf{immersed in the experience}. As one participant noted: ``\textit{It’s about allowing yourself to fully experience it, feeling completely immersed and satisfied. You can also explain why it’s called a ‘feeling’—because the emotions are heightened to the fullest.}'' (P6, follow-up interview) Through this immersive activity, the participants gradually began to reconnect with their bodies, allowing themselves to feel the stress and tension that had been building up. %This process enabled them to become more aware of the bodily sensations they typically overlooked or ignored in their daily lives.

\subsubsection{Feel and Connect to the Stressed Bodies though Somatosensory}
\hh{Our findings revealed that many participants \textbf{rarely engage with their bodily sensations} in everyday life, often due to the pressure of intense academic routines. As one participant shared: \textit{“You rarely take the time to sit quietly and observe your body.” }(P4, follow-up interview)}. Through somatosensory methods such as body scanning, the participants began to become more \textbf{aware of various bodily experiences}, including sensations of pain, numbness, and comfort. The participants noted that body scanning give them the opportunities to notice their bodies. One participant reported the sensation of having a shackle around her neck, as her neck felt immobile when the body scanning facilitator guided the scan through that area, indicating the tension caused by stress. (P1, body mapping exercise) \hh{These moments of embodied awareness served as a foundation for later self-disclosure, helping participants articulate how stress was experienced and held in the body.}

% The connection between body and mind is perhaps one of the most obvious yet most overlooked facts for many people. Chronic stress leads people to habitually ignore their bodies and emotions \cite{mariotti_effects_2015} and this was acknowledged by our participants:\textit{“You rarely take the time to sit quietly and observe your body.” } (P4, follow-up interview)  The participants \textbf{seldom have the opportunity to feel their bodies}, as the pressures of their intense academic lives leave little room for such awareness. Their focus is often on studies, leaving bodily sensations and their well-being overlooked. 

\subsubsection{Self-disclose the Bodily Sensations of Stress through Visual and Tangible Ways}
\hh{Our findings show that embodied methods such as body mapping and tangible sculpting supported participants in translating their bodily sensations of stress into \textbf{visual and material forms}. This process of externalising internal experiences enabled participants to move beyond purely verbal modes of self-expression and provided alternative channels for self-disclosure. As P5 described:} \textit{“Being able to visualise what was observed and present it in front of me felt quite magical. It felt like moving from a sensory experience to something visual, something tangible.”} \hh{Our analysis revealed that these alternative modes of expression supported self-disclosure in several interconnected ways.}

% \hh{By supporting this translation from felt sensation to material representation, the embodied methods played a key role in enabling self-disclosure, which were further explained in the follows}
%In the follow-up interview, participants explained how they valued using embodied methods such as body mapping for translating their personal observations and abstract bodily feelings into \textbf{visual or tangible representations}: 

%This enabled participants to visualise and verbalise stress, which is something normally intangible and difficult to articulate. By providing them with opportunities to more fully and precisely express their sensations, participants are better able to articulate and understand the stress stored in their bodies.

\hh{First, the creation of these artefacts served as a form of embodied reflection, allowing participants to engage more deeply with their lived experiences of stress. As they worked on their body maps, they began to interpret bodily sensations in relation to personal memories or habitual conditions.} \textit{“Since I’ve been doing homework for a long time and my posture hasn’t been very good, I developed a ‘buffalo hump’\footnote{A buffalo hump is a buildup of fat in the upper back and neck, causing a hump-like shape.}. My buffalo hump has always been painful, and as she (body scanning facilitator) was talking about it, I could feel the pain even more.” }(P9, body mapping exercise) In such moments, the representations of bodily sensations became a prompt for further verbal and emotional articulation. 

\begin{figure*}[h]
    \centering
    \includegraphics[width=1\linewidth]{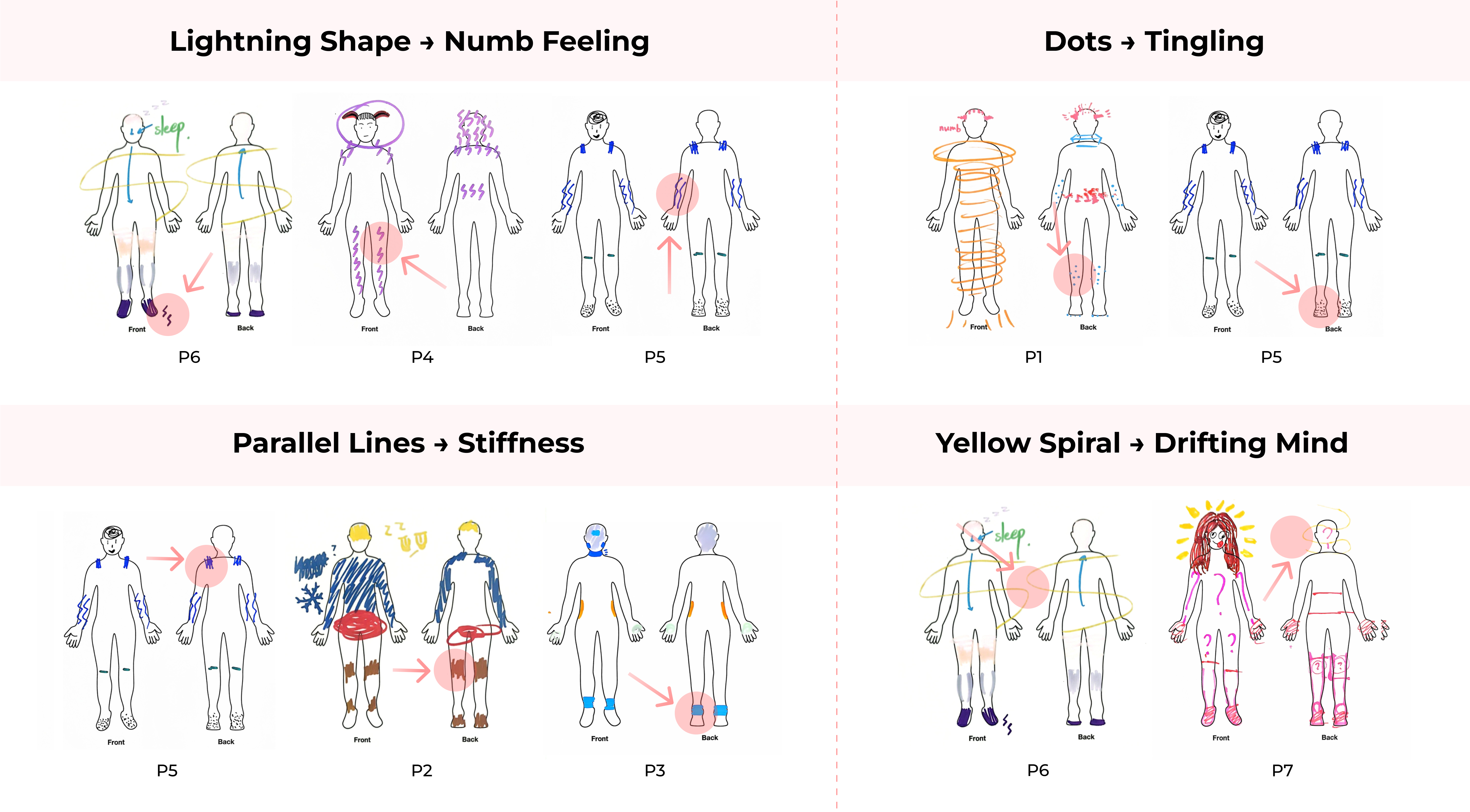}
    \caption{Similar metaphorical imageries identified on body maps.}
    \Description{Fig. 3: A composite figure divided into four quadrants, each illustrating participants’ bodily sensations linked to specific visual patterns and mental or emotional states. Top left shows “Lightning Shape → Numb Feeling” with body outlines marked by jagged lines on limbs. Top right shows “Dots → Tingling” with dotted patterns across torsos and legs. Bottom left is “Parallel Lines → Stiffness” with straight line patterns across shoulders, thighs, and legs. Bottom right shows “Yellow Spiral → Drifting Mind” featuring spirals and abstract shapes around the torso and head. Each quadrant includes front and back outlines of the human body marked with colourful annotations and participant IDs (P1–P7) to indicate individual contributions.}
    \label{fig4}
\end{figure*}

\begin{figure}[h]
  \centering
  \begin{minipage}{0.45\textwidth}
    \centering
    \includegraphics[width=\textwidth]{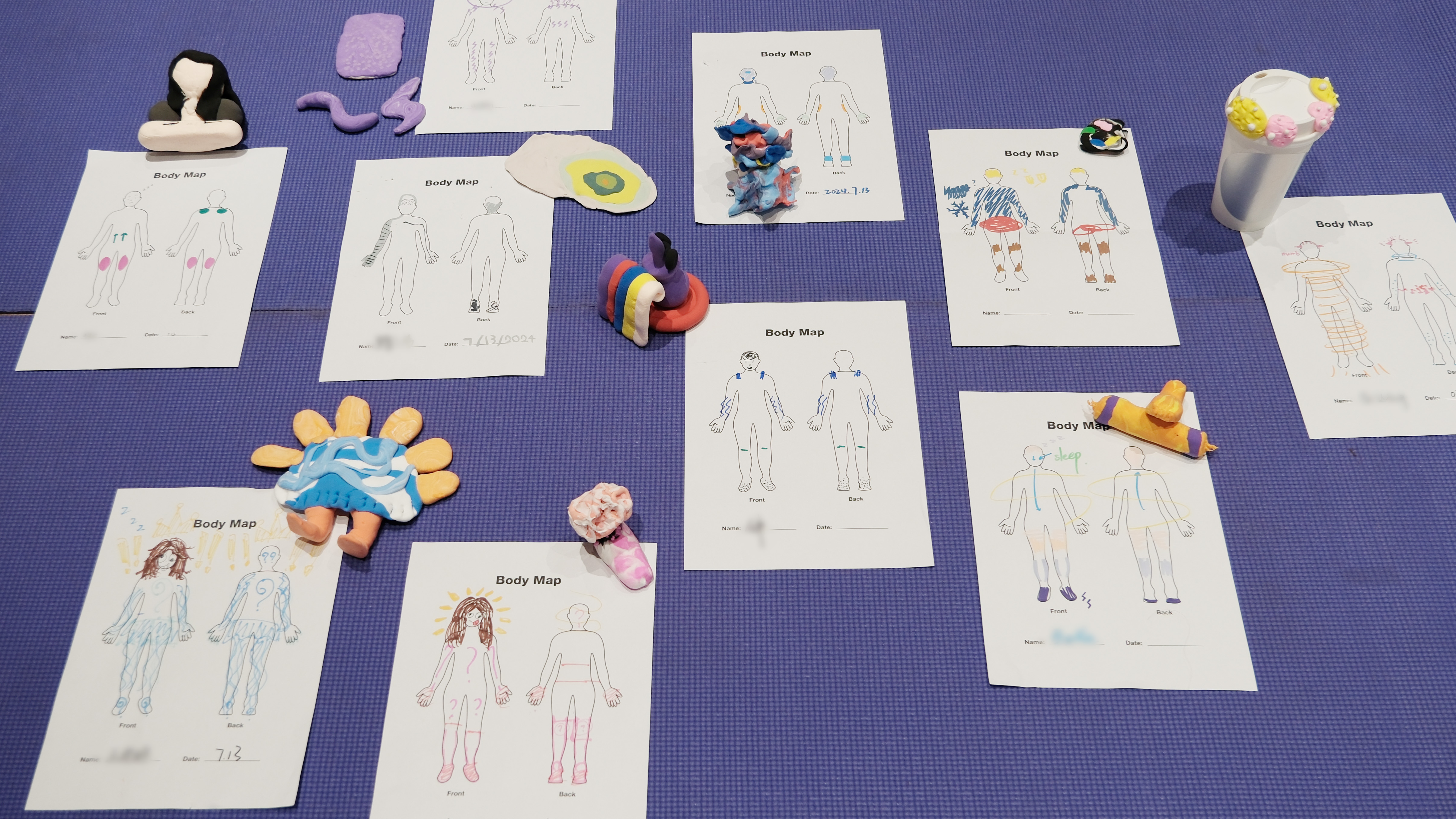}
    \caption{Body maps and tangible body maps sculpture.}
    \Description{Fig. 4: A photo showing ten illustrated body maps created by participants, each paired with a matching tangible body mapping sculpture made from soft clay materials.}
    \label{fig5}
  \end{minipage}
  \hspace{0.5cm} % 控制图像之间的间距
  \begin{minipage}{0.45\textwidth}
    \centering
    \includegraphics[width=\textwidth]{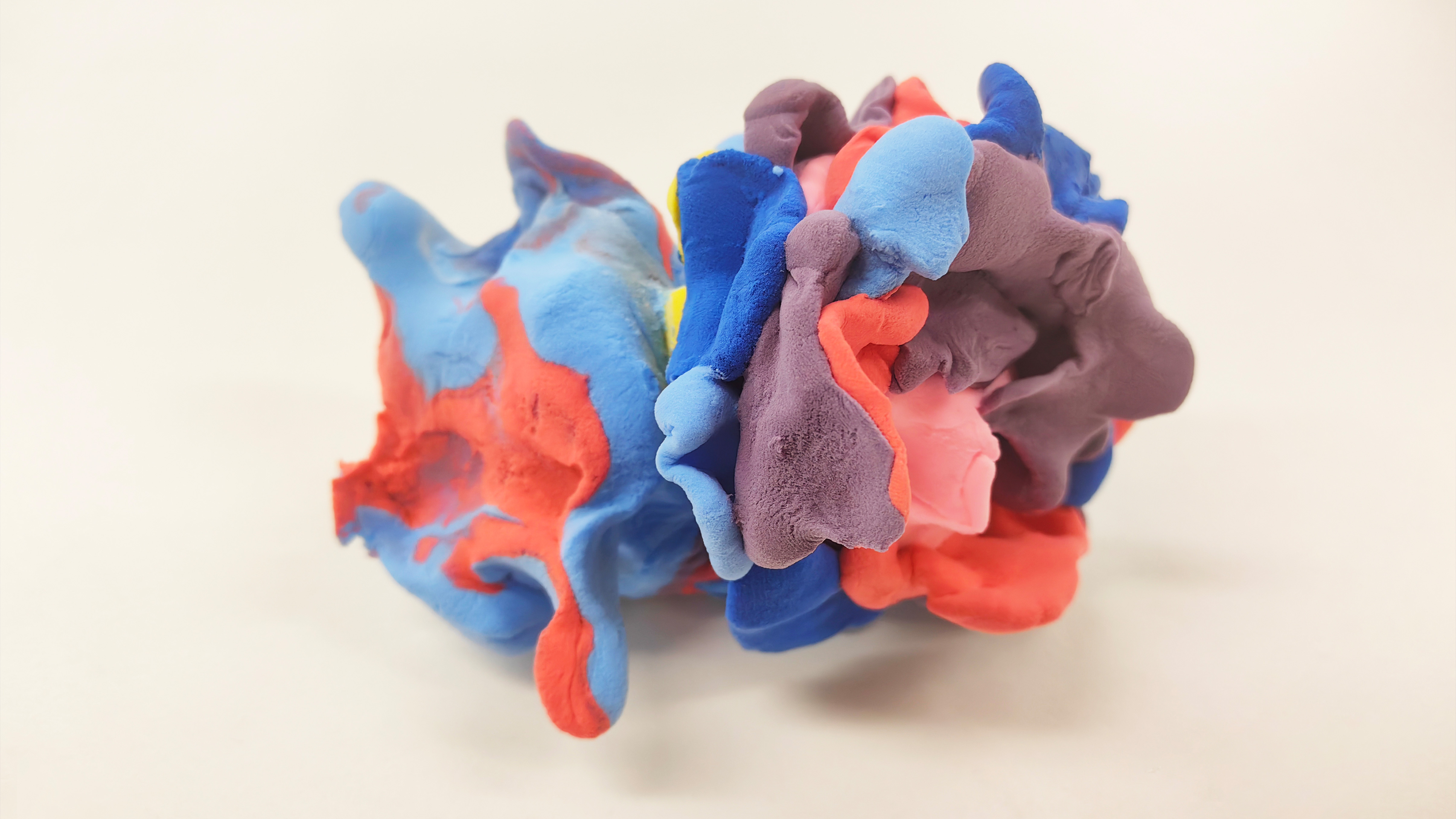}
    \caption{Tangible body map sculpture by P3.}
    \Description{Fig. 5: A photo of a tangible body map clay sculpture featuring organic, petal-like forms in red, blue, and brown hues.}
    \label{fig6}
  \end{minipage}
\end{figure}

\hh{Additionally, we noticed that participants frequently drew on \textbf{metaphorical imagery} to express their embodied sensations, using recurring visual elements to represent difficult-to-describe feelings.} For example, with the body maps, three participant used parallel and consecutive lines to illustrate the feeling of stiffness, while three others used lightning shapes to manifest the numb feeling (\autoref{fig4}). Two participants used dots to convey a tingling sensation during the workshop: \textit{“My whole foot just feels like it has ants crawling on it.”} (P5, body mapping exercise) These metaphors were also present in the tangible body mapping activity. In the sculptures created for tangible body maps (\autoref{fig3}, S2), the representation involved greater emphasis on form and materiality compared to traditional body maps (\autoref{fig5}). For instance, P3 shaped a statue of herself to express the emptiness of her mind: \textit{“[…]it feels like mushrooms growing on my head, and then underneath it has eight legs, and I felt like I’m turning into an octopus.”} (P3, tangible body map exercise) Metaphors like mushrooms growing and turning into an octopus exhibit strong embodied qualities (\autoref{fig6}). 

%In the sculptures created for tangible body maps (\autoref{fig3}, S2), the representation involved greater emphasis on form and materiality compared to body maps (\autoref{fig5}). P3 shaped a statue of herself to express the emptiness of her mind: \textit{“[…]it feels like mushrooms growing on my head, and then underneath it has eight legs, and I felt like I’m turning into an octopus.”} (P3, tangible body map exercise) Metaphors like mushrooms growing and turning into an octopus exhibit strong embodied qualities (\autoref{fig6}). These tangible representations exhibit a deeper level of awareness of stress.

% The method of tangible body mapping was then used for the participants to shape a certain experience of stress, which helped them further reflect on their lived experience.\textit{“There are two lumps here, and if my shoulders are sore, it feels like there are stones pressing down on them, which is very uncomfortable.”} (P9, tangible body mapping exercise) Through these embodied methods, the participants noticed more details of their bodily experience, allowing them to better recognise their stress and, as a result, engage in more effective self-disclosure.

%The participants expressed themselves through drawing and sculpture, followed by verbal expression. As they spoke, they used metaphors, linking their experiences to the embodied methods. The patterns that represented the bodily sensations of stress on both the body maps and tangible body maps consistently showed similarities (\autoref{fig4}). We refer to these patterns as \textbf{embodied metaphors} \cite{dauden_roquet_body_2020}. Embodied metaphors were frequently used by the participants when they needed to translate their bodily experience into physical representations of stress. 

\subsubsection{Leverage the Body to Intervene Disclosed Stress}
\begin{figure*}[h]
    \centering
    \includegraphics[width=1\linewidth]{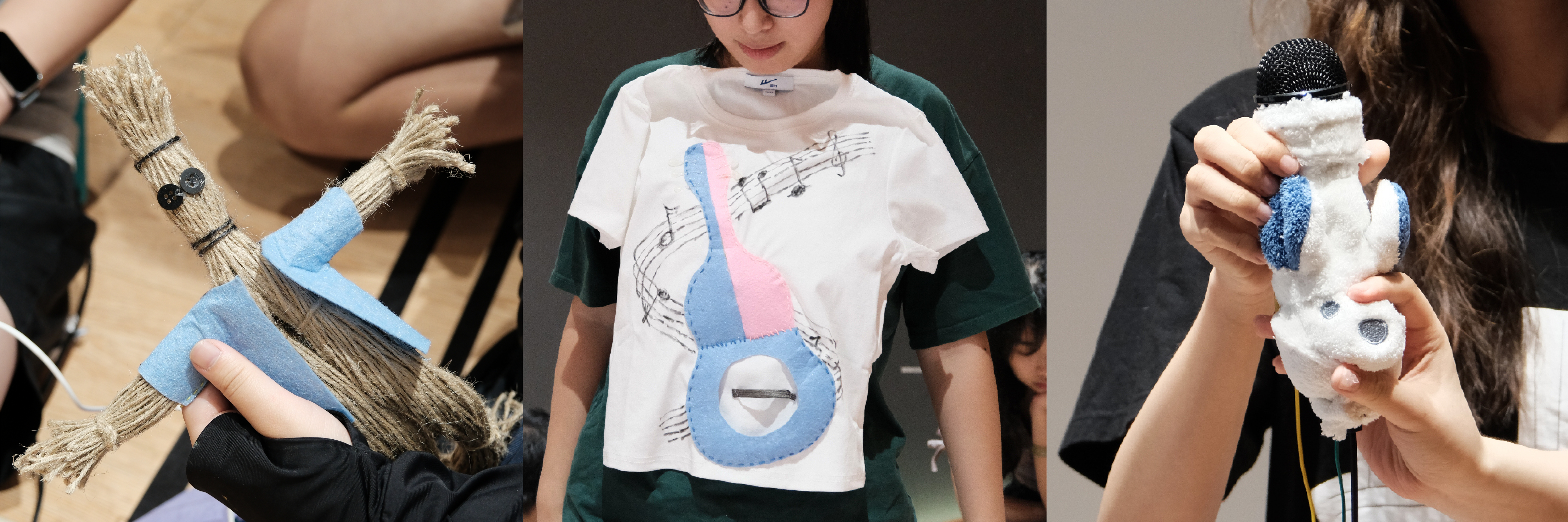}
    \caption{Three working prototypes created in workshops to cope with stress:  (left) “Hell Girl”, a stress-relieving toy for emotional venting through interactive squeezing by P1-3, (centre) “Riffing T-shirt”,  a wearable guitar for stress relief through auditory interaction by P4-6, (right) “EmoMic”, an interactive microphone for stress relief through playful voice modulation by P7-9.}
    \Description{Fig. 6: An image displaying three interactive prototypes developed by participants during the workshops: “Hell Girl” (left), a stress-relieving toy for emotional venting through interactive squeezing by P1-3, “Riffing T-shirt” (centre), a wearable guitar for stress relief through auditory interaction by P4-6, “EmoMic” (right), an interactive microphone for stress relief through playful voice modulation by P7-9.}
    \label{fig7}
\end{figure*}

\xl{Participants used their \textbf{bodies as a medium} to express and share stress experiences, which supported the creation of three wearable stress-relief prototypes. Each design was grounded in a disclosed stressor and aimed to amplify a specific pleasant bodily sensation as a form of emotional support. The embodied making process with the EmTex toolkit helped participants externalise internal stress through physical forms, making it easier to share personal experiences and reflect on them together.} Two prototypes, “Hell Girl” (\autoref{fig7}, left) and “EmoMic” (\autoref{fig7}, right) linked the action of squeezing to sound generation, which participants identified as the most stress-relieving behaviour. For a full description of the prototype, please refer to the appendix. 

At the same time, we also found that when participants acted as the designers, they had more opportunities to disclose about their stress because they need to explain their ideas or justify their design decisions to others. For example, one team designed a T-shirt that could be played as a guitar (\autoref{fig7}, middle). P5 explained their insights towards stress-relief: \textit{“We designed this wearable to help release stress. We noticed that when people feel stressed, they often turn to auditory experiences for comfort and healing. So, we decided to combine auditory elements with clothing to create an integrated solution.”} We found that during the ideation and fabrication process of the prototypes, the participants constantly disclosed more about stress when they work with their teammates. These hands-on activities used making as an expressive tool to facilitate self-disclosure.

In summary, our embodied approaches to stress self-disclosure provide a holistic way to communicate about stress, bridging the gap between mind and body. These methods help individuals, especially adolescents who constantly had their feelings ignored by themselves, to externalise and make sense of their lived experience of stress, offering a powerful approach for expressing and coping stress. 

\subsection{Self-disclosed dimensions of Stress (RQ2)}
\xl{Through an embodied approach, adolescent girls disclosed different dimensions of stress, including somatic symptoms of stress, sources of stress and means of relieving stress. Our analysis shows the unique characteristics of adolescent girls' stress and the potential of these insights to inform interventions for stress-related mental well-being. Below, we present our detailed findings.}

\subsubsection{Somatic Symptom of Stress}
The first dimension of stress self-disclosure that could be self-disclosed via embodied approach is somatic symptom of stress. The most commonly experienced somatic symptom of stress disclosed by our participants was pain. A total count of  21 cases of pain were identified across our dataset. These included \textbf{chronic pain} such as lower back, shoulder and neck pain. Three participants reported that they suffered from multiple kinds of pain when facing stress: \textit{“I have migraines, normally on the left side. Sometimes my whole face aches. […] My stomach aches before exams, it is a kind of twisted pain.”} (P6, focus group session). 

\textbf{Gastrointestinal symptoms}, such as stomach pain, loss of appetite, and nausea, were also experienced by several participants. \textit{“I had a psychological consultation, and the therapist told me that when people are under a lot of stress, their first reaction is often physical, particularly in the stomach. For example, they might lose their appetite, and in more severe cases, they may feel like vomiting.” }(P3,  focus group session)  Two other participants acknowledged these gastrointestinal symptoms.

\textbf{Musculoskeletal symptoms} such as stiffness in the shoulders and neck was also commonly reported as somatic symptoms together with numb body ends (\autoref{fig4}, upper-left corner). Other reported somatic symptoms included cramps, menstrual disorders and proprioceptive dysfunction in the context of stress: \textit{“When I feel my knees, it feels like my two knees are not the same length; they're just not in the right place, they're not symmetrical.”} (P7, body mapping exercise). We found that stress not only triggered somatic symptoms but also made participants feel less in control of their bodies,as these physical manifestations are closely tied to emotional distress and affect how adolescents perceive and relate to their own bodies.

%During the workshops, we observed that when describing their somatic symptoms, most spoke as if they were explaining their conditions to a doctor and used medical terminology. \textit{“I've had chronic functional dyspepsia since my second year of high school, which I believe was triggered by high levels of stress. […] I often experience discomfort in the epigastric region, feeling very bloated and uncomfortable after every meal.”} (P12, identify stress symptoms exercise) Other participants also used medical terminology such as “palpitations” and “rheumatism”. We also noticed that when our young participants discussed their stressed bodies, they often objectified them, referring to their bodies from a third-person perspective and disconnecting from them: \textit{“\textbf{It's }around the cervical spine. If I look at the computer for too long or carry something too heavy,\textbf{ it }becomes really painful. Also, when my hands are especially tense, they shake a lot.”} (P14, identify stress symptoms exercise)

\subsubsection{Sources of Stress}
The second dimension of stress self-disclosure concerns the sources of stress. \textbf{Academic stress} was the most frequently mentioned, with participants also revealing peer pressure, parental expectations and self-esteem issues under this theme. One participant described her embodied experience of feeling sick whilst taking a math exam: \textit{“When I got to question 12, I thought that question was so disgusting, and said to myself, how could anyone come up with a question like that, […]Then I started to feel like vomiting, and my whole body was very dizzy.”}(P9, focus group session) P9’s experience illustrates how her entire body was engaged, communicating the embodied quality of the lived experience of stress. When recounting her experience, P9 also went through an embodied interpretation process: she described a feeling of disgust towards the exam question, which was transformed into actual physical discomfort. Another interesting finding from the participants’ stress self-disclosure is that their own expectations were mentioned more frequently than parental expectations as a source of stress. \textit{“The less hope your family has for you, the more pressure you feel, like a rebellious urge. You want to prove them wrong, and in doing so, you place pressure on yourself. This kind of pressure, where there’s no direct pressure from others but it feels overwhelming, is actually the most devastating.”} (P4, focus group session) Participants tended to put pressure on themselves instead of receiving pressures from their parents, as parental expectations can be implicit and have been internalised by participants. 

%\rh{<- the quotes in this paragraph are great! As they represent the UNIQUE findings of our study: \textbf{linking stressor with their bodily experience}. Maybe we should better leverage these quotes in the follow paragraph when describing academic stress.}

%\rh{->} After adopting embodied approach such as body scanning and body mapping, the participants were in a better position to explore and reflect their sources of stress: \textit{“You rarely have the opportunity to open up and share your thoughts. It’s mostly been about studying, studying, and more studying, with hardly any chance to truly connect with your own feelings.”} (P5, follow-up interview) That’s why the Embodied Probes workshop was designed to create a space that didn’t feel like a classroom, but instead fostered a safe and relaxed environment for the participants. \rh{<- overlap with RQ1 findings, as you are presenting ``effectiveness of our workshop'', not the stress dimensions of AD. } 

\textbf{Interpersonal stress} as a source of stress was coded 11 times in our database.  Several participants shared a common fear of being excluded by their peers, which increased their interpersonal stress. This is illustrated next in an academic context where the pressure from peer relationships made the situation more intricate: \textit{“Someone who works hard alone might be isolated. For example, when I study vocabulary by myself, people start talking about me, mocking me, saying I'm ‘overcompetitive’ and that I'm pressuring them to compete as well.”} (P9, focus group session) 

Participants also disclosed several other significant sources of stress including stereotypical view, body image, and family dynamics. Several participants shared that they experienced \textbf{stereotype threat} in the school context associated with underperformance and under representation, a psychological phenomenon that has been documented before in mathematics and science fields \cite{peck_inducing_2020}. For example, when we first approached the school with our workshop proposal, the teachers asked if we intended to involve the boys, as they believed the girls were not good at programming and that the boys could provide assistance. In the Embodied Probes workshop, participants also reported that one of their teachers holds stereotypical view towards girls,  which was summarised as followed : \textit{“If you outperform boys in exams, it's because you're lucky; if you don't, it's because you deserve it.”} (P4, follow-up interview) We found that such stereotypical views diminished girls' agency, which in turn affected their stress levels.

The participants are also under societal pressures regarding appearance. They are worried if they fail to meet social standards of beauty, which is influenced by mass media and pop culture. In the tangible body mapping exercise, one participant stated that she had her focus on her hip: \textit{“Because I felt that my focus was on my hips the most, I shaped it into something like a skirt.”} (P8, tangible body map exercise) The embodied methods later helped her to disclose her stress towards her \textbf{body image}:\textit{“I experience body image anxiety because some of my friends are very slim and don't have a little belly. But I feel like my stomach is large, and every day I tell myself I need to lose weight. However, in reality, I can't seem to take action, so I still feel a lot of pressure.”} (P8, follow-up interview) In our workshops, we found that participants tended to make comparison with their peers, leading to dissatisfaction of their own body image, which in turn increased their anxiety. 

%\rh{<- These sentences are great! As they give people the feeling that you are mainly describing the stressor. though you mentioned some embodied methods, they serve as the context information, not the protagonist. Thumbs! You may consider the nuanced differences in writing and apply such style in your future writings. } 

Some participants who have younger siblings also reported uncomfortable \textbf{family dynamics} as a source of stress. Their parents expected them to meet the needs of younger siblings: \textit{“Right now, I am teaching him to study, and in the future, I may have to support his education.”} (P4, focus group session) Although later the participant clarified that her parents might not be serious about this, she still felt obligated to meet the family expectations. Other participants explained that their parents expected them, as the older child, to serve as a role model for their younger siblings.: \textit{“There is a younger one at home, my little sister, so I definitely have to set a good example for her because my sister doesn't like studying. That's why my mom wants me to study well to show her.”} (P5, focus group session) Overall, participants reported their feelings of having to take on educational responsibilities, which contributed to their stress.

\subsubsection{Means of Stress Relief}
The last dimension of stress self-disclosure focuses on stress relief. Here we focus on the non-verbal means of stress relief. \xl{We focus on the means that are closely related to embodied experience, emphasising the means that reflect how stress is felt and expressed through the body.} Haptic qualities of \textbf{relaxation objects} were highly valued for their ability to support stress relief. For example, several participants introduced their “Abebe” that help them reduce stress. Abebe is a recent Chinese internet meme referring to objects that provide emotional support and have accompanied someone for a long time. P5 gave an example of her Abebe: \textit{“I have a Stitch plushie that I've been hugging since the second year of middle school. Its stuffing has been flattened from all the hugging, and its neck is even broken. Yet, I still need to hold it to sleep.”} (P5, focus group session) P8 added: \textit{“I think having a pet for companionship is also wonderful. A puppy or anything fluffy would do.”} (P8, focus group session) 

\xl{One modality our participants emphasised for stress relief was tactile sensation. \textit{“I especially love large dogs because when petting them, the sense of touch feels particularly comforting.”} (P8, focus group session) Another \textbf{sensory modality} identified by the participants which supports stress relieving and relaxation was smell}: \textit{“The smell of sun-dried dust mites on quilts is the most stress-relieving.”} (P4, focus group session) She later develop this notion of using smell to relieve stress in brainstorming with her teammates: \textit{“When experiencing high levels of stress, smelling a pleasant fragrance can provide a calming and soothing effect. Therefore, we hope to create a device that releases a soothing scent when it detects elevated stress levels, helping individuals alleviate their stress.”} (P4-6, brainstorming session) 

Several participants also mentioned that engaging in \textbf{creative activities} with their body was stress-relieving.  For them, the process of making and creating not only brought enjoyment but also served as an effective way to manage the stress encountered in their daily live. A few participants reflected on their making process, pointing out that it was the immersive quality, which made handcrafting stress-relieving. \textit{“It’s my flow moment.”} (P5, follow-up interview)  The creative activities encouraged self-disclosure by prompting adolescent girls to reflect on why and how they wanted to address their stressed bodies, thereby further facilitating their self-disclosure. \xl{Many of the design ideas generated in the workshop were inspired by direct somatic reflections. One design solution (see Appendix \autoref{fig8}, "Hell Girl"), for instance, compromised a scarecrow toy that could be squeezed in moments of anxiety, drawing directly from participants' habitual gestures of clenching fists under stress.} Some participants also described a great sense of achievement upon completing a piece of handcraft. \textit{“That's right, I feel a great sense of accomplishment.”} (P7, follow-up interview)  Overall, engaging in crafting activities in the workshop not only provided a means of stress relief but also served to empower adolescent girls by fostering creativity, building confidence, and offering a sense of accomplishment.

\section{Discussion}
\label{discussion}
\hh{While much research on self-disclosure has focused on verbal modalities, stress is often experienced and held in the body. This makes embodied approaches not only relevant but essential for understanding and supporting stress-related self-expression. In this paper, we conducted two workshops and analysed the findings to explore how the body, as a medium, can facilitate stress-related self-disclosure among adolescent girls. In the following discussion, we reflect on our approach by: 1) examining how the embodied approach unfolded in our study, 2) offering implications and encouraging future research to support adolescent girls' well-being, and 3) acknowledging the limitations of our work.}

%While much research has focused on verbal modalities of self-disclosure, stress is deeply connected to our bodies, making embodied approach to self-disclosure an under explored area. In this paper, we conducted two workshops and analysed the findings to explore how the body as a medium can facilitate stress self-disclosure. In the discussion, we reflect on our approach by addressing: 1) how the embodied approach unfolds, and \xl{2) how future research can build on these unique insights to generate design implications, particularly for supporting adolescent girls in subsequent studies.}

\subsection{Taking an Embodied Approach for Stress Self-Disclosure}
 \xl{Previous research on stress self-disclosure has largely overlooked the role of the body. Given the close relationship between stress and bodily experiences, our study explored how the body can serve as a medium for stress self-disclosure. To investigate this, we conducted a series of workshops incorporating a range of embodied methods. Our first key finding (RQ1) illustrates how these methods supported and facilitated stress self-disclosure among adolescent girls.}
 
When discussing stress, people often associate it primarily with the mind, viewing stress through a body-mind dualism lens \cite{vitetta_mindbody_2005}. Stress is typically regarded as intangible and abstract. However, through the lens of embodiment \cite{dourish_where_2001}, we were able to uncover the bodily dimension of stress. \xl{Following typical soma design practices, we crafted a series of workshop activities that centred on bodily experience, enabling adolescent girls to actively engage with and reflect on their felt sensations. These embodied activities supported participants in becoming more aware of how stress manifests in their bodies, allowing them to surface and articulate bodily dimensions of stress that are often overlooked.} Feeling the body is fundamental to understanding and connecting with oneself.\textbf{ }The method we used is body scanning \cite{sauer2013comparing}. We hope to\textbf{ inform future interactive technologies to assist people experience more nuanced and detailed bodily feelings.} Currently, most body scanning activities involve passive listening, but with interactive devices, individuals might be able to experience different bodily sensations. For example, we envision wearable technologies like smart clothing or textile-based sensors could provide real-time feedback on muscle tension or stress, offering tactile sensations (e.g., vibrations or warmth) that help people better understand their body’s responses.

\xl{In our study, bodily experience served as the foundation for stress self-disclosure. Unlike prior work that primarily aimed to deepen somatic awareness or enhance user experience, our goal was to support self-disclosure. While methods such as body scanning effectively facilitated rich bodily sensations, it was a challenge to guide adolescent girls with no prior exposure to soma design to externalise what they felt in their bodies.} Besides encouraging verbal sharing of stress, we found the body mapping method \cite{anne_cochrane_body_2022} to be effective in helping participants visually represent their bodily experiences. Building on previous work by Núñez-Pacheco on tangible body mapping \cite{nunez-pacheco_dialoguing_2022}, we guided participants to focus on a specific bodily sensation to deepen their awareness of stress in a tangible way. Embodied methods provided the participants more channels to communicate clearly what kind of bodily feelings one is experiencing, helped participants articulate a range of bodily-related somatic symptoms and lived experiences associated with stress. \xl{Building on our findings, future research could further explore \textbf{how to support body-based self-disclosure through the design of interactive and storytelling-based tools that help individuals externalise and verbalise their stress experiences.} Drawing inspiration from play therapy \cite{ray_use_2013}, where therapists observe how children interact with objects in a sandbox rather than relying solely on verbal communication, HCI researchers could similarly investigate more tangible and embodied approaches to facilitate the expression of stress-related experiences.}
%Drawing inspiration from play therapy \cite{ray_use_2013}, where therapists observe how patients interact with objects in the sandbox rather than relying solely on direct communication, HCI researchers, much like therapists, could \textbf{explore more tangible ways to better support individuals in expressing their bodily feelings and engaging self-disclosure.}

\xl{During the body mapping process, we observed an interesting phenomenon, the use of metaphorical imagery.} We refer to these patterns as embodied metaphors \cite{dauden_roquet_body_2020}, which allowing deeper level of self-awareness and expression, not just through words, but through form, volume, and imaginative narrative. \xl{Participants described their bodily sensations using vivid metaphors, such as comparing the feeling in their feet to ants crawling or likening the sensation in their heads to mushrooms growing. These metaphors not only enriched the expression of somatic experiences but also served as an accessible and imaginative way for the girls to externalise and communicate sensations that might otherwise be difficult to articulate. While previous HCI research has explored the use of metaphors to represent various experiences, such as learning processes \cite{reed2023negotiating} or cognitive states \cite{behfar2023can}, our study reveals how metaphors can specifically be employed to connect bodily sensations. This insight opens up new possibilities for future soma design research, which could explore\textbf{ how metaphors not only help individuals articulate and relate to their own bodily experiences but also foster empathy, enabling them to connect with and understand others’ bodily sensations.}}

\xl{In addition to body mapping, we also incorporated expressive making activities into our study. Traditionally, in soma design procedures, design activities are primarily aimed at generating concepts or prototypes. However, our study highlights how such activities can also function as a form of self-disclosure. Their designs themselves can also be seen as a form of intervention, suggesting a future research direction where adolescents create personalised interventions based on their own embodied experiences. The process of justifying their design decisions became an avenue for further self-expression and reflection. These design solutions emerged from a process where participants were encouraged to discover how stress manifested in their bodies. These somatic insights helped participants articulate their psychological needs that might not have surfaced through conventional verbal discussion alone. In this way, the workshop outcomes affirmed the value of embodied approach in helping adolescents externalise and design around abstract, often unspoken, lived experiences of stress. We suggest that future research could explore\textbf{ how to deliberately structure design activities as effective means for facilitating self-disclosure, particularly in contexts involving sensitive, embodied experiences.}}

\xl{This study extends soma design into the context of stress self-disclosure, a domain where the connection between stress and bodily sensation renders embodied approaches particularly relevant. In this context, soma design is not only a method of bodily exploration but also a means to support self-disclosure. By grounding expression in somatic awareness, participants are enabled to externalise and communicate stress through tangible, embodied forms. This approach highlights the potential of soma design to scaffold more intentional and reflective stress self-disclosure practices. Furthermore, we argue that the act of materialising stress, rendering it visible and shareable, can itself function as a form of intervention, pointing to future directions for soma-based research and design that centres on body-informed strategies for emotional well-being.}

\subsection{Design Implications for Adolescent Girls' Stress}
\xl{Through the embodied methods and processes described above, our Findings 2 (RQ2) surfaced a rich set of empirical insights related to the stress experiences of adolescent girls. Our study was initially motivated by the lack of attention to this demographic within HCI research on stress, and we see these findings as an important contribution toward addressing this gap. Reflecting on the specific stress-related experiences we uncovered, we propose several key design implications that warrant further attention in future research.}

\xl{By focusing on the embodied experiences of adolescent girls, our study reveals that stress-related somatic symptoms such as chronic pain, gastrointestinal discomfort and musculoskeletal issues have received limited attention in prior HCI research.} While many HCI studies have focused on biofeedback systems to detect and manage stress \cite{yu_biofeedback_2018, umair_thermopixels_2020}, somatic symptoms of stress have received relatively little attention. We argue that\textbf{ the somatic symptoms disclosed by users are a crucial dimension of stress.} Therefore, future research in stress management should further explore these embodied aspects and their potential role in effective stress interventions. We also note that while some of these somatic symptoms have been examined in relation to older populations, such as chronic pain among elderly users \cite{rapo2016chronic}, they are also relevant to adolescent girls. These symptoms are closely linked to the physical and emotional transitions of adolescence, including hormonal changes, growth-related discomfort and heightened sensitivity to body image \cite{forney2019interaction}. Given these distinctive factors, we suggest that future HCI research should\textbf{ further investigate how stress is physically experienced by younger populations.}

%The lived experience of pain, for example, can only be fully understood through the lens of embodiment, as the pain is not just an abstract concept but something that is felt deeply physical \cite{svenaeus_phenomenology_2015}. 

Based on our findings, we identified that the sources of stress for Chinese adolescent girls can be categorised into academic, interpersonal, and sociocultural stress. Within sociocultural stress, our participants disclosed experiences of stereotype threat and body image anxiety. \xl{One important aspect is the role of stereotypes, which has already received attention in prior HCI research \cite{peck_inducing_2020}.} Societal norms and narratives around women come with predefined expectations, placing unique pressures and demands on adolescent girls \cite{keleher2008changing}. These \xl{stressors} can sometimes \xl{cause} negative impacts on adolescent girls. For example, girls are more likely to develop eating disorders than boys \cite{noauthor_mental_2017}. \xl{Moreover, Chinese adolescent girls experience significant academic stress\cite{sun2012academic}, further complicating their overall well-being and placing them in a more vulnerable position.} \xl{We suggest that further research in HCI should\textbf{ }\textbf{attend to the specific mental health needs of adolescent girls, addressing the gendered and demographic dimensions of stress through more inclusive design approaches.} Our findings fill a gap in HCI by shedding light on the experiences of Chinese adolescent girls. While these stressors may also appear in other cultures, their forms and impacts can differ, highlighting the need for further cross-cultural research.}

\xl{Our findings on stress relief primarily focused on non-verbal strategies, such as the use of relaxation objects, engagement with sensory modalities, and participation in creative activities. HCI has explored a range of non-verbal interactions related to emotion and stress, such as affective computing \cite{picard2000affective}, tangible interfaces \cite{alonso2008squeeze}, and olfactory interactions \cite{amores2018bioessence}.} One interesting finding was the participants found the sense of smell stress relieving. Smell has been linked to memory in psychological studies, with autobiographical memories shown to be evoked by sensory cues \cite{willander_smell_2006}. \xl{A promising direction for future work is to explore\textbf{ how these non-verbal approaches can be thoughtfully adapted to the unique needs and lived experiences of adolescent girls.}} In our study, we also discovered that \xl{adolescent girls} found creative activities effective for stress relief. They expressed significant enthusiasm for the making process and highlighted the sense of accomplishment they felt upon completing the \xl{prototype}. Creativity and empowerment shares a complementary relationship \cite{velthouse1990creativity}. \xl{ In the process of designing these solutions, participants not only explored and expressed stress in a more holistic and tangible manner but also gained a sense of empowerment. These findings suggest that future research could further investigate \textbf{how creative activities can be structured to promote both emotional well-being and empowerment among adolescent girls}.}

\xl{Finally, we reflect on our workshop design. We made a number of deliberate design considerations in structuring the workshops to specifically support the needs and characteristics of this demographic and may also offer valuable insights for future research seeking to engage adolescent girls in stress-related discussions.} Intimate self-disclosure can have potential benefits for mental well-being, but it can also be harmful if done in an insecure environment \cite{soleymani_multimodal_2019}.  \xl{To avoid triggering academic stress associated with traditional classroom settings, the self-disclosure activities were conducted in a lecture theatre space that was intentionally arranged to foster a more relaxed atmosphere.} Working with one’s body can also make people vulnerable and additional considerations were needed when planning the workshop \cite{hook_designing_2018}. Previous research showed that the pressure of the male gaze can constrain women’s personal development and may limit their ability or confidence to seek support on personal matters \cite{naseem_designing_2020}. Therefore, limiting participation in the workshop to girls only helped fostering a safe environment and in our case, participants positively commented on the absence of boys as reported in Section 6.1.1. Here we build on previous work on intimate health \cite{sondergaard_designing_2021-1, sondergaard_designing_2020, campo_woytuk_touching_2020} and offering a new case study in adolescent girls’ mental well-being. \xl{Future research could explore\textbf{ how physical space, such as lighting and seating, shapes participants’ sense of safety, and examine how group dynamics and social factors, like gender composition, influence participants' comfort and openness.}}

Furthermore, we adapted a youth-centred approach and considered \xl{adolescent girls} to be our co-designers/co-researchers. Co-design methods have been widely used by researchers with patients, children, and other vulnerable groups \cite{bate_experience-based_2006, price_childrens_2024}. Participants generated stress-relief ideas from their bodily experience shared at the workshop and took part in hands-on fabrication activities to realise their designs. Through these creative activities, the participants put themselves in the position of a designer. In the process they need to justify their decisions, which also act as a form of self-disclosure when solving problems. This approach encouraged \xl{adolescent girls} to open up more with their peers and achieved greater self-disclosure at the same time.  Including young people as co-researchers on a subject matter that is important to them has proven successful in previous research practices \cite{lygnegaard2023adolescent}. \xl{In the focus group session, participants took on the role of researchers, a shift that encouraged self-disclosure and enabled them to gather valuable insights for this study.} \textbf{Involving adolescents as design/research partners can help HCI researchers empathise with their user group whilst developing bespoke designs that have the potential to inform future interactive technologies that address youth-centred needs. }

\xl{This research foregrounds the specific experiences of adolescent girls in China, a group uniquely positioned at the intersection of intense academic pressure, evolving bodily awareness, and sociocultural expectations. By focusing on this demographic, the study surfaces distinctive insights, such as the overlooked somatic dimensions of stress, culturally and developmentally specific stressors, and the potential of non-verbal and creative strategies. These insights highlight the need for design approaches that are both culturally sensitive and developmentally appropriate. The workshop design was shaped by careful design considerations, such as creating a non-threatening environment, avoiding potential body image triggers, and fostering empowerment among participants with creative activities.}

\subsection{Limitations and Future Work}
While the Embodied Probes was successful in the workshop settings, limitations such as the limited participant source (all from schools in Shanghai) and the small sample size need to be addressed in future research. Shanghai is an economically advanced city with a relatively more open mindset compared to other regions in China. Therefore, the self-disclosures made by the participants in the workshop may not necessarily reflect those of girls across the entire country. \xl{Considering that the participants are partially familiar with each other, the dynamics of whether they fully know or do not know each other may affect the workshop outcomes, and this requires further exploration. Another limitation is that the screening of participants' mental status was based solely on self-reports, even though they were confirmed by parents or teachers; future research should pay particular attention to potential body image and eating disorders.}

As an exploratory study, we tried mainstream embodied methods such as body scanning and body mapping. While alternative approaches may lead to different interpretations, we argue that the overall direction of the embodied approach is valuable. Future research could explore additional methods in this area. We aim to further develop the Embodied Probes into a comprehensive toolkit that not only facilitates stress self-disclosure but also supports the broader mental well-being goals of adolescent girls. Our intention is to deploy this toolkit with the same group of adolescents over an extended period to gain deeper insights into their mental well-being needs. Additionally, we aim to explore how Embodied Probes can support the mental well-being of young people in diverse cultural contexts. To this end, we are initiating a comparative study between China and the UK, which we hope will provide more diverse insights related to stress and help us further understand how we can adapt embodied methods and design interactive technologies for stress self-disclosure in different contexts. 
\section{Conclusion}
In conclusion, this study highlights the potential of Embodied Probes as tools for stress self-disclosure among adolescent girls through a two-days workshop with nine participants. Our findings indicate that embodied methods facilitate stress self-disclosure across multiple dimensions, such as somatic symptoms of stress, sources of stress and means of stress relief. Additionally, co-designed interactive interventions can act as empowering tools for young women in managing stress. This research contributes to the HCI community by introducing a novel embodied approach to stress self-disclosure, providing design implications that can inform future interactive self-disclosure technologies and further enrich the intersection of HCI and women’s mental well-being.

\begin{acks}
\xl{We extend our heartfelt thanks to the participants in these workshops and sincerely acknowledge their valuable contribution to this research. We are also thankful to the anonymous ACs and reviewers for their generous and constructive feedback on this paper. We thank Tongji University for approving the ethical review of this project, under the approval number: tjdxsr2024062. This research was funded by National Natural Science Foundation of China (project number 62202335), "Sino-German Cooperation 2.0" Funding Program of Tongji University (project number ZD2023029), and Innovative Design and Intelligent Manufacturing (IDIM) Discipline Group Project. Caroline Claisse’s work on this paper was covered by the UKRI-funded project the Centre for Digital Citizens (project number EP/T022582/1).}
\end{acks}

\bibliographystyle{ACM-Reference-Format}
\balance
\bibliography{sample-base}


\begin{thebibliography}{79}

%%% the Web address.

\ifx \showCODEN    \undefined \def \showCODEN     #1{\unskip}     \fi
\ifx \showISBNx    \undefined \def \showISBNx     #1{\unskip}     \fi
\ifx \showISBNxiii \undefined \def \showISBNxiii  #1{\unskip}     \fi
\ifx \showISSN     \undefined \def \showISSN      #1{\unskip}     \fi
\ifx \showLCCN     \undefined \def \showLCCN      #1{\unskip}     \fi
\ifx \shownote     \undefined \def \shownote      #1{#1}          \fi
\ifx \showarticletitle \undefined \def \showarticletitle #1{#1}   \fi
\ifx \showURL      \undefined \def \showURL       {\relax}        \fi
% invisible to TeX
\providecommand\bibfield[2]{#2}
\providecommand\bibinfo[2]{#2}
\providecommand\natexlab[1]{#1}
\providecommand\showeprint[2][]{arXiv:#2}

\bibitem[noa(2017)]%
        {noauthor_mental_2017}
 \bibinfo{year}{2017}\natexlab{}.
\newblock \showarticletitle{Mental {Health} of {Children} and {Young} {People} in {England}, 2017 [{PAS}] - {NHS} {England} {Digital}}.
\newblock  (\bibinfo{year}{2017}).
\newblock


\bibitem[Alonso et~al\mbox{.}(2008)]%
        {alonso2008squeeze}
\bibfield{author}{\bibinfo{person}{Miguel~Bruns Alonso}, \bibinfo{person}{David~V Keyson}, {and} \bibinfo{person}{Caroline~CM Hummels}.} \bibinfo{year}{2008}\natexlab{}.
\newblock \showarticletitle{Squeeze, rock, and roll; can tangible interaction with affective products support stress reduction?}. In \bibinfo{booktitle}{\emph{Proceedings of the 2nd international conference on Tangible and embedded interaction}}. \bibinfo{pages}{105--108}.
\newblock


\bibitem[Amores et~al\mbox{.}(2018)]%
        {amores2018bioessence}
\bibfield{author}{\bibinfo{person}{Judith Amores}, \bibinfo{person}{Javier Hernandez}, \bibinfo{person}{Artem Dementyev}, \bibinfo{person}{Xiqing Wang}, {and} \bibinfo{person}{Pattie Maes}.} \bibinfo{year}{2018}\natexlab{}.
\newblock \showarticletitle{Bioessence: A wearable olfactory display that monitors cardio-respiratory information to support mental wellbeing}. In \bibinfo{booktitle}{\emph{2018 40th Annual International Conference of the IEEE Engineering in Medicine and Biology Society (EMBC)}}. IEEE, \bibinfo{pages}{5131--5134}.
\newblock


\bibitem[Andalibi(2020)]%
        {andalibi_disclosure_2020}
\bibfield{author}{\bibinfo{person}{Nazanin Andalibi}.} \bibinfo{year}{2020}\natexlab{}.
\newblock \showarticletitle{Disclosure, {Privacy}, and {Stigma} on {Social} {Media}: {Examining} {Non}-{Disclosure} of {Distressing} {Experiences}}.
\newblock \bibinfo{journal}{\emph{ACM Transactions on Computer-Human Interaction}} \bibinfo{volume}{27}, \bibinfo{number}{3} (\bibinfo{date}{June} \bibinfo{year}{2020}), \bibinfo{pages}{1--43}.
\newblock
\showISSN{1073-0516, 1557-7325}
\href{https://doi.org/10.1145/3386600}{doi:\nolinkurl{10.1145/3386600}}
\newblock
\shownote{Publisher: Association for Computing Machinery (ACM)}.


\bibitem[Anne~Cochrane et~al\mbox{.}(2022)]%
        {anne_cochrane_body_2022}
\bibfield{author}{\bibinfo{person}{Karen Anne~Cochrane}, \bibinfo{person}{Kristina Mah}, \bibinfo{person}{Anna Ståhl}, \bibinfo{person}{Claudia Núñez-Pacheco}, \bibinfo{person}{Madeline Balaam}, \bibinfo{person}{Naseem Ahmadpour}, {and} \bibinfo{person}{Lian Loke}.} \bibinfo{year}{2022}\natexlab{}.
\newblock \showarticletitle{Body {Maps}: {A} {Generative} {Tool} for {Soma}-based {Design}}. In \bibinfo{booktitle}{\emph{Sixteenth {International} {Conference} on {Tangible}, {Embedded}, and {Embodied} {Interaction}}}. \bibinfo{publisher}{ACM}, \bibinfo{address}{Daejeon Republic of Korea}, \bibinfo{pages}{1--14}.
\newblock
\showISBNx{978-1-4503-9147-4}
\href{https://doi.org/10.1145/3490149.3502262}{doi:\nolinkurl{10.1145/3490149.3502262}}


\bibitem[Bate and Robert(2006)]%
        {bate_experience-based_2006}
\bibfield{author}{\bibinfo{person}{P. Bate} {and} \bibinfo{person}{G. Robert}.} \bibinfo{year}{2006}\natexlab{}.
\newblock \showarticletitle{Experience-based design: from redesigning the system around the patient to co-designing services with the patient}.
\newblock \bibinfo{journal}{\emph{Quality and Safety in Health Care}} \bibinfo{volume}{15}, \bibinfo{number}{5} (\bibinfo{date}{Oct.} \bibinfo{year}{2006}), \bibinfo{pages}{307--310}.
\newblock
\showISSN{1475-3898, 1475-3901}
\href{https://doi.org/10.1136/qshc.2005.016527}{doi:\nolinkurl{10.1136/qshc.2005.016527}}


\bibitem[Behfar et~al\mbox{.}(2023)]%
        {behfar2023can}
\bibfield{author}{\bibinfo{person}{Arezou Behfar}, \bibinfo{person}{Hanieh Atashpanjeh}, {and} \bibinfo{person}{Mahdi~Nasrullah Al-Ameen}.} \bibinfo{year}{2023}\natexlab{}.
\newblock \showarticletitle{Can Password Meter be More Effective Towards User Attention, Engagement, and Attachment?: A Study of Metaphor-based Designs}. In \bibinfo{booktitle}{\emph{Companion Publication of the 2023 Conference on Computer Supported Cooperative Work and Social Computing}}. \bibinfo{pages}{164--171}.
\newblock


\bibitem[Braun and Clarke(2006)]%
        {braun_using_2006}
\bibfield{author}{\bibinfo{person}{Virginia Braun} {and} \bibinfo{person}{Victoria Clarke}.} \bibinfo{year}{2006}\natexlab{}.
\newblock \showarticletitle{Using thematic analysis in psychology}.
\newblock \bibinfo{journal}{\emph{Qualitative Research in Psychology}} \bibinfo{volume}{3}, \bibinfo{number}{2} (\bibinfo{date}{Jan.} \bibinfo{year}{2006}), \bibinfo{pages}{77--101}.
\newblock
\showISSN{1478-0887, 1478-0895}
\href{https://doi.org/10.1191/1478088706qp063oa}{doi:\nolinkurl{10.1191/1478088706qp063oa}}


\bibitem[Campo~Woytuk et~al\mbox{.}(2020)]%
        {campo_woytuk_touching_2020}
\bibfield{author}{\bibinfo{person}{Nadia Campo~Woytuk}, \bibinfo{person}{Marie Louise~Juul Søndergaard}, \bibinfo{person}{Marianela Ciolfi~Felice}, {and} \bibinfo{person}{Madeline Balaam}.} \bibinfo{year}{2020}\natexlab{}.
\newblock \showarticletitle{Touching and {Being} in {Touch} with the {Menstruating} {Body}}. In \bibinfo{booktitle}{\emph{Proceedings of the 2020 {CHI} {Conference} on {Human} {Factors} in {Computing} {Systems}}}. \bibinfo{publisher}{ACM}, \bibinfo{address}{Honolulu HI USA}, \bibinfo{pages}{1--14}.
\newblock
\showISBNx{978-1-4503-6708-0}
\href{https://doi.org/10.1145/3313831.3376471}{doi:\nolinkurl{10.1145/3313831.3376471}}


\bibitem[Catalan et~al\mbox{.}(2024)]%
        {catalan_storytelling_2024}
\bibfield{author}{\bibinfo{person}{Cristobal Catalan}, \bibinfo{person}{Lina Gega}, {and} \bibinfo{person}{Jonathan Hook}.} \bibinfo{year}{2024}\natexlab{}.
\newblock \showarticletitle{Storytelling {Games} for {General} {Anxiety}: {Clinician} {Perspectives} on {Walking} {Simulator} {Games} as {Intervention}}.
\newblock \bibinfo{journal}{\emph{Proceedings of the ACM on Human-Computer Interaction}} \bibinfo{volume}{8}, \bibinfo{number}{CHI PLAY} (\bibinfo{date}{Oct.} \bibinfo{year}{2024}), \bibinfo{pages}{1--24}.
\newblock
\showISSN{2573-0142}
\href{https://doi.org/10.1145/3677104}{doi:\nolinkurl{10.1145/3677104}}


\bibitem[Claisse et~al\mbox{.}(2022)]%
        {claisse_tangible_2022}
\bibfield{author}{\bibinfo{person}{Caroline Claisse}, \bibinfo{person}{Muhammad Umair}, \bibinfo{person}{Abigail~C Durrant}, \bibinfo{person}{Charles Windlin}, \bibinfo{person}{Pavel Karpashevich}, \bibinfo{person}{Kristina Höök}, \bibinfo{person}{Vasiliki Tsaknaki}, \bibinfo{person}{Pedro Sanches}, {and} \bibinfo{person}{Corina Sas}.} \bibinfo{year}{2022}\natexlab{}.
\newblock \showarticletitle{Tangible {Interaction} for {Supporting} {Well}-being}. In \bibinfo{booktitle}{\emph{{CHI} {Conference} on {Human} {Factors} in {Computing} {Systems} {Extended} {Abstracts}}}. \bibinfo{publisher}{ACM}, \bibinfo{address}{New Orleans LA USA}, \bibinfo{pages}{1--5}.
\newblock
\showISBNx{978-1-4503-9156-6}
\href{https://doi.org/10.1145/3491101.3503716}{doi:\nolinkurl{10.1145/3491101.3503716}}


\bibitem[Cohen(1988)]%
        {cohen_perceived_1988}
\bibfield{author}{\bibinfo{person}{Sheldon Cohen}.} \bibinfo{year}{1988}\natexlab{}.
\newblock \showarticletitle{Perceived stress in a probability sample of the {United} {States}.}
\newblock In \bibinfo{booktitle}{\emph{The social psychology of health.}} \bibinfo{publisher}{Sage Publications, Inc}, \bibinfo{address}{Thousand Oaks, CA, US}, \bibinfo{pages}{31--67}.
\newblock
\showISBNx{0-8039-3162-X (Hardcover); 0-8039-3163-8 (Paperback)}


\bibitem[Compas et~al\mbox{.}(1993)]%
        {compas1993adolescent}
\bibfield{author}{\bibinfo{person}{Bruce~E Compas}, \bibinfo{person}{Pamela~G Orosan}, {and} \bibinfo{person}{Kathryn~E Grant}.} \bibinfo{year}{1993}\natexlab{}.
\newblock \showarticletitle{Adolescent stress and coping: Implications for psychopathology during adolescence}.
\newblock \bibinfo{journal}{\emph{Journal of adolescence}} \bibinfo{volume}{16}, \bibinfo{number}{3} (\bibinfo{year}{1993}), \bibinfo{pages}{331--349}.
\newblock


\bibitem[Cozby(1973)]%
        {cozby_self-disclosure_1973}
\bibfield{author}{\bibinfo{person}{Paul~C. Cozby}.} \bibinfo{year}{1973}\natexlab{}.
\newblock \showarticletitle{Self-disclosure: {A} literature review.}
\newblock \bibinfo{journal}{\emph{Psychological Bulletin}} \bibinfo{volume}{79}, \bibinfo{number}{2} (\bibinfo{year}{1973}), \bibinfo{pages}{73--91}.
\newblock
\showISSN{1939-1455, 0033-2909}
\href{https://doi.org/10.1037/h0033950}{doi:\nolinkurl{10.1037/h0033950}}


\bibitem[Daudén~Roquet and Sas(2020)]%
        {dauden_roquet_body_2020}
\bibfield{author}{\bibinfo{person}{Claudia Daudén~Roquet} {and} \bibinfo{person}{Corina Sas}.} \bibinfo{year}{2020}\natexlab{}.
\newblock \showarticletitle{Body {Matters}: {Exploration} of the {Human} {Body} as a {Resource} for the {Design} of {Technologies} for {Meditation}}. In \bibinfo{booktitle}{\emph{Proceedings of the 2020 {ACM} {Designing} {Interactive} {Systems} {Conference}}}. \bibinfo{publisher}{ACM}, \bibinfo{address}{Eindhoven Netherlands}.
\newblock
\href{https://doi.org/10.1145/3357236.3395499}{doi:\nolinkurl{10.1145/3357236.3395499}}


\bibitem[Dourish(2001)]%
        {dourish_where_2001}
\bibfield{author}{\bibinfo{person}{Paul Dourish}.} \bibinfo{year}{2001}\natexlab{}.
\newblock \bibinfo{booktitle}{\emph{Where the action is: the foundations of embodied interaction}}.
\newblock \bibinfo{publisher}{MIT Press}, \bibinfo{address}{Cambridge, Mass}.
\newblock
\showISBNx{978-0-262-04196-6}


\bibitem[Edman et~al\mbox{.}(2017)]%
        {edman2017perceived}
\bibfield{author}{\bibinfo{person}{Joel~S Edman}, \bibinfo{person}{Jeffrey~M Greeson}, \bibinfo{person}{Rhonda~S Roberts}, \bibinfo{person}{Adam~B Kaufman}, \bibinfo{person}{Donald~I Abrams}, \bibinfo{person}{Rowena~J Dolor}, {and} \bibinfo{person}{Ruth~Q Wolever}.} \bibinfo{year}{2017}\natexlab{}.
\newblock \showarticletitle{Perceived stress in patients with common gastrointestinal disorders: associations with quality of life, symptoms and disease management}.
\newblock \bibinfo{journal}{\emph{Explore}} \bibinfo{volume}{13}, \bibinfo{number}{2} (\bibinfo{year}{2017}), \bibinfo{pages}{124--128}.
\newblock


\bibitem[Farber(2006)]%
        {farber_self-disclosure_2006}
\bibfield{author}{\bibinfo{person}{Barry~A. Farber}.} \bibinfo{year}{2006}\natexlab{}.
\newblock \bibinfo{booktitle}{\emph{Self-disclosure in psychotherapy}}.
\newblock \bibinfo{publisher}{Guilford Press}, \bibinfo{address}{New York}.
\newblock
\showISBNx{978-1-59385-323-5}
\newblock
\shownote{OCLC: ocm65617438}.


\bibitem[Fiorilli et~al\mbox{.}(2019)]%
        {fiorilli2019predicting}
\bibfield{author}{\bibinfo{person}{Caterina Fiorilli}, \bibinfo{person}{Teresa Grimaldi~Capitello}, \bibinfo{person}{Daniela Barni}, \bibinfo{person}{Ilaria Buonomo}, {and} \bibinfo{person}{Simonetta Gentile}.} \bibinfo{year}{2019}\natexlab{}.
\newblock \showarticletitle{Predicting adolescent depression: The interrelated roles of self-esteem and interpersonal stressors}.
\newblock \bibinfo{journal}{\emph{Frontiers in psychology}}  \bibinfo{volume}{10} (\bibinfo{year}{2019}), \bibinfo{pages}{565}.
\newblock


\bibitem[Forney et~al\mbox{.}(2019)]%
        {forney2019interaction}
\bibfield{author}{\bibinfo{person}{K~Jean Forney}, \bibinfo{person}{Pamela~K Keel}, \bibinfo{person}{Shannon O'Connor}, \bibinfo{person}{Cheryl Sisk}, \bibinfo{person}{S~Alexandra Burt}, {and} \bibinfo{person}{Kelly~L Klump}.} \bibinfo{year}{2019}\natexlab{}.
\newblock \showarticletitle{Interaction of hormonal and social environments in understanding body image concerns in adolescent girls}.
\newblock \bibinfo{journal}{\emph{Journal of Psychiatric Research}}  \bibinfo{volume}{109} (\bibinfo{year}{2019}), \bibinfo{pages}{178--184}.
\newblock


\bibitem[Foster et~al\mbox{.}(2025)]%
        {foster2025embodiment}
\bibfield{author}{\bibinfo{person}{Lo Foster}, \bibinfo{person}{Lars-Gunnar Lundh}, {and} \bibinfo{person}{Daiva Daukantait{\.e}}.} \bibinfo{year}{2025}\natexlab{}.
\newblock \showarticletitle{Embodiment and Psychological Health in Adolescence: 1. Development and Validation of a Brief 12-item Questionnaireto Measure the Experience of Embodiment}.
\newblock \bibinfo{journal}{\emph{Journal for Person-Oriented Research}} \bibinfo{volume}{11}, \bibinfo{number}{1} (\bibinfo{year}{2025}), \bibinfo{pages}{10}.
\newblock


\bibitem[Foundation(2024)]%
        {noauthor_waht_nodate}
\bibfield{author}{\bibinfo{person}{Interaction~Design Foundation}.} \bibinfo{year}{2024}\natexlab{}.
\newblock \bibinfo{title}{What is {Brainstorming}? 10 {Effective} {Techniques} {You} {Can} {Use}}.
\newblock
\urldef\tempurl%
\url{https://www.interaction-design.org/literature/topics/brainstorming}
\showURL{%
\tempurl}
\newblock
\shownote{Accessed: 2024-09-28}.


\bibitem[Francis(2018)]%
        {francis_embodied_2018}
\bibfield{author}{\bibinfo{person}{Alisha~L. Francis}.} \bibinfo{year}{2018}\natexlab{}.
\newblock \showarticletitle{The {Embodied} {Theory} of {Stress}: {A} {Constructionist} {Perspective} on the {Experience} of {Stress}}.
\newblock \bibinfo{journal}{\emph{Review of General Psychology}} \bibinfo{volume}{22}, \bibinfo{number}{4} (\bibinfo{date}{Dec.} \bibinfo{year}{2018}), \bibinfo{pages}{398--405}.
\newblock
\showISSN{1089-2680, 1939-1552}
\href{https://doi.org/10.1037/gpr0000164}{doi:\nolinkurl{10.1037/gpr0000164}}
\newblock
\shownote{Publisher: SAGE Publications}.


\bibitem[Gaver et~al\mbox{.}(1999)]%
        {gaver_design_1999}
\bibfield{author}{\bibinfo{person}{Bill Gaver}, \bibinfo{person}{Tony Dunne}, {and} \bibinfo{person}{Elena Pacenti}.} \bibinfo{year}{1999}\natexlab{}.
\newblock \showarticletitle{Design: {Cultural} probes}.
\newblock \bibinfo{journal}{\emph{Interactions}} \bibinfo{volume}{6}, \bibinfo{number}{1} (\bibinfo{date}{Jan.} \bibinfo{year}{1999}), \bibinfo{pages}{21--29}.
\newblock
\showISSN{1072-5520, 1558-3449}
\href{https://doi.org/10.1145/291224.291235}{doi:\nolinkurl{10.1145/291224.291235}}


\bibitem[Glise et~al\mbox{.}(2014)]%
        {glise_prevalence_2014}
\bibfield{author}{\bibinfo{person}{Kristina Glise}, \bibinfo{person}{Gunnar Ahlborg}, {and} \bibinfo{person}{Ingibjörg~H Jonsdottir}.} \bibinfo{year}{2014}\natexlab{}.
\newblock \showarticletitle{Prevalence and course of somatic symptoms in patients with stress-related exhaustion: does sex or age matter}.
\newblock \bibinfo{journal}{\emph{BMC Psychiatry}} \bibinfo{volume}{14}, \bibinfo{number}{1} (\bibinfo{date}{Dec.} \bibinfo{year}{2014}), \bibinfo{pages}{118}.
\newblock
\showISSN{1471-244X}
\href{https://doi.org/10.1186/1471-244X-14-118}{doi:\nolinkurl{10.1186/1471-244X-14-118}}


\bibitem[Gonsalves et~al\mbox{.}(2023)]%
        {gonsalves_systematic_2023}
\bibfield{author}{\bibinfo{person}{Pattie~P. Gonsalves}, \bibinfo{person}{Rithika Nair}, \bibinfo{person}{Madhavi Roy}, \bibinfo{person}{Sweta Pal}, {and} \bibinfo{person}{Daniel Michelson}.} \bibinfo{year}{2023}\natexlab{}.
\newblock \showarticletitle{A {Systematic} {Review} and {Lived} {Experience} {Synthesis} of {Self}-disclosure as an {Active} {Ingredient} in {Interventions} for {Adolescents} and {Young} {Adults} with {Anxiety} and {Depression}}.
\newblock \bibinfo{journal}{\emph{Administration and Policy in Mental Health and Mental Health Services Research}} \bibinfo{volume}{50}, \bibinfo{number}{3} (\bibinfo{date}{May} \bibinfo{year}{2023}), \bibinfo{pages}{488--505}.
\newblock
\showISSN{0894-587X, 1573-3289}
\href{https://doi.org/10.1007/s10488-023-01253-2}{doi:\nolinkurl{10.1007/s10488-023-01253-2}}
\newblock
\shownote{Publisher: Springer Science and Business Media LLC}.


\bibitem[Gustafsson et~al\mbox{.}(2009)]%
        {gustafsson2009perceived}
\bibfield{author}{\bibinfo{person}{Sanna~Aila Gustafsson}, \bibinfo{person}{Birgitta Edlund}, \bibinfo{person}{Josefine Dav{\'e}n}, \bibinfo{person}{Lars Kjellin}, {and} \bibinfo{person}{Claes Norring}.} \bibinfo{year}{2009}\natexlab{}.
\newblock \showarticletitle{Perceived expectations in daily life among adolescent girls suffering from an eating disorder: A phenomenographic study}.
\newblock \bibinfo{journal}{\emph{Eating Disorders}} \bibinfo{volume}{18}, \bibinfo{number}{1} (\bibinfo{year}{2009}), \bibinfo{pages}{25--42}.
\newblock


\bibitem[Haraldsson et~al\mbox{.}(2011)]%
        {haraldsson_adolescent_2011}
\bibfield{author}{\bibinfo{person}{Katarina Haraldsson}, \bibinfo{person}{Eva‐Carin Lindgren}, \bibinfo{person}{Bengt Mattsson}, \bibinfo{person}{Bengt Fridlund}, {and} \bibinfo{person}{Bertil Marklund}.} \bibinfo{year}{2011}\natexlab{}.
\newblock \showarticletitle{Adolescent girls' experiences of underlying social processes triggering stress in their everyday life: a grounded theory study}.
\newblock \bibinfo{journal}{\emph{Stress and Health}} \bibinfo{volume}{27}, \bibinfo{number}{2} (\bibinfo{date}{April} \bibinfo{year}{2011}).
\newblock
\showISSN{1532-3005, 1532-2998}
\href{https://doi.org/10.1002/smi.1336}{doi:\nolinkurl{10.1002/smi.1336}}


\bibitem[Höök(2018)]%
        {hook_designing_2018}
\bibfield{author}{\bibinfo{person}{Kristina Höök}.} \bibinfo{year}{2018}\natexlab{}.
\newblock \bibinfo{booktitle}{\emph{Designing with the {Body}: {Somaesthetic} {Interaction} {Design}}}.
\newblock \bibinfo{publisher}{The MIT Press}, \bibinfo{address}{Cambridge, Massachusetts London}.
\newblock


\bibitem[Höök et~al\mbox{.}(2021)]%
        {hook_unpacking_2021}
\bibfield{author}{\bibinfo{person}{Kristina Höök}, \bibinfo{person}{Steve Benford}, \bibinfo{person}{Paul Tennent}, \bibinfo{person}{Vasiliki Tsaknaki}, \bibinfo{person}{Miquel Alfaras}, \bibinfo{person}{Juan~Martinez Avila}, \bibinfo{person}{Christine Li}, \bibinfo{person}{Joseph Marshall}, \bibinfo{person}{Claudia~Daudén Roquet}, \bibinfo{person}{Pedro Sanches}, \bibinfo{person}{Anna Ståhl}, \bibinfo{person}{Muhammad Umair}, \bibinfo{person}{Charles Windlin}, {and} \bibinfo{person}{Feng Zhou}.} \bibinfo{year}{2021}\natexlab{}.
\newblock \showarticletitle{Unpacking {Non}-{Dualistic} {Design}: {The} {Soma} {Design} {Case}}.
\newblock \bibinfo{journal}{\emph{ACM Transactions on Computer-Human Interaction}} \bibinfo{volume}{28}, \bibinfo{number}{6} (\bibinfo{date}{Dec.} \bibinfo{year}{2021}), \bibinfo{pages}{1--36}.
\newblock
\showISSN{1073-0516, 1557-7325}
\href{https://doi.org/10.1145/3462448}{doi:\nolinkurl{10.1145/3462448}}


\bibitem[Jung and Stolterman(2010)]%
        {jung_material_2010}
\bibfield{author}{\bibinfo{person}{Heekyoung Jung} {and} \bibinfo{person}{Erik Stolterman}.} \bibinfo{year}{2010}\natexlab{}.
\newblock \showarticletitle{Material probe: exploring materiality of digital artifacts}. In \bibinfo{booktitle}{\emph{Proceedings of the fifth international conference on {Tangible}, embedded, and embodied interaction}}. \bibinfo{publisher}{ACM}, \bibinfo{address}{Funchal Portugal}, \bibinfo{pages}{153--156}.
\newblock
\showISBNx{978-1-4503-0478-8}
\href{https://doi.org/10.1145/1935701.1935731}{doi:\nolinkurl{10.1145/1935701.1935731}}


\bibitem[Jusoh et~al\mbox{.}(2024)]%
        {jusoh_helpbot_2024}
\bibfield{author}{\bibinfo{person}{Shaidah Jusoh}, \bibinfo{person}{Hejab Al~Fawareh}, \bibinfo{person}{Rabiah Abdul~Kadir}, {and} \bibinfo{person}{Hassan Hosseinzadeh}.} \bibinfo{year}{2024}\natexlab{}.
\newblock \showarticletitle{{HelpBot}: {A} {Web}-{Based} {Chatbot} to {Handle} {Depression} {Among} {Adolescents}}. In \bibinfo{booktitle}{\emph{Proceedings of the 2024 10th {International} {Conference} on {Computing} and {Artificial} {Intelligence}}}. \bibinfo{publisher}{ACM}, \bibinfo{address}{Bali Island Indonesia}, \bibinfo{pages}{149--154}.
\newblock
\showISBNx{9798400717055}
\href{https://doi.org/10.1145/3669754.3669777}{doi:\nolinkurl{10.1145/3669754.3669777}}


\bibitem[Keleher and Franklin(2008)]%
        {keleher2008changing}
\bibfield{author}{\bibinfo{person}{Helen Keleher} {and} \bibinfo{person}{Lucinda Franklin}.} \bibinfo{year}{2008}\natexlab{}.
\newblock \showarticletitle{Changing gendered norms about women and girls at the level of household and community: a review of the evidence}.
\newblock \bibinfo{journal}{\emph{Global public health}} \bibinfo{volume}{3}, \bibinfo{number}{S1} (\bibinfo{year}{2008}), \bibinfo{pages}{42--57}.
\newblock


\bibitem[Kitson et~al\mbox{.}(2024)]%
        {kitson_i_2024}
\bibfield{author}{\bibinfo{person}{Alexandra Kitson}, \bibinfo{person}{Alissa~N. Antle}, \bibinfo{person}{Sadhbh Kenny}, \bibinfo{person}{Ashu Adhikari}, \bibinfo{person}{Kenneth Karthik}, \bibinfo{person}{Artun Cimensel}, {and} \bibinfo{person}{Melissa Chan}.} \bibinfo{year}{2024}\natexlab{}.
\newblock \showarticletitle{'{I} {Call} {Upon} a {Friend}': {Virtual} {Reality}-{Based} {Supports} for {Cognitive} {Reappraisal} {Identified} through {Co}-designing with {Adolescents}}. In \bibinfo{booktitle}{\emph{Proceedings of the {CHI} {Conference} on {Human} {Factors} in {Computing} {Systems}}}. \bibinfo{publisher}{ACM}, \bibinfo{address}{Honolulu HI USA}, \bibinfo{pages}{1--18}.
\newblock
\showISBNx{9798400703300}
\href{https://doi.org/10.1145/3613904.3642723}{doi:\nolinkurl{10.1145/3613904.3642723}}


\bibitem[Lau et~al\mbox{.}(2016)]%
        {lau2016adolescents}
\bibfield{author}{\bibinfo{person}{Anna~S Lau}, \bibinfo{person}{Sisi Guo}, \bibinfo{person}{William Tsai}, \bibinfo{person}{D~Julie Nguyen}, \bibinfo{person}{Hannah~T Nguyen}, \bibinfo{person}{Victoria Ngo}, {and} \bibinfo{person}{Bahr Weiss}.} \bibinfo{year}{2016}\natexlab{}.
\newblock \showarticletitle{Adolescents’ stigma attitudes toward internalizing and externalizing disorders: Cultural influences and implications for distress manifestations}.
\newblock \bibinfo{journal}{\emph{Clinical psychological science}} \bibinfo{volume}{4}, \bibinfo{number}{4} (\bibinfo{year}{2016}), \bibinfo{pages}{704--717}.
\newblock


\bibitem[Li et~al\mbox{.}(2023)]%
        {li_tell_2023}
\bibfield{author}{\bibinfo{person}{Tony~W. Li}, \bibinfo{person}{Michael Murray}, \bibinfo{person}{Zander Brumbaugh}, \bibinfo{person}{Raida Karim}, \bibinfo{person}{Hanna Lee}, \bibinfo{person}{Maya Cakmak}, {and} \bibinfo{person}{Elin~A. Björling}.} \bibinfo{year}{2023}\natexlab{}.
\newblock \showarticletitle{Tell {Me} {About} {It}: {Adolescent} {Self}-{Disclosure} with an {Online} {Robot} for {Mental} {Health}}. In \bibinfo{booktitle}{\emph{Companion of the 2023 {ACM}/{IEEE} {International} {Conference} on {Human}-{Robot} {Interaction}}}, Vol.~\bibinfo{volume}{525}. \bibinfo{publisher}{ACM}, \bibinfo{address}{Stockholm Sweden}, \bibinfo{pages}{183--187}.
\newblock
\href{https://doi.org/10.1145/3568294.3580068}{doi:\nolinkurl{10.1145/3568294.3580068}}


\bibitem[Loke and Schiphorst(2018)]%
        {loke_somatic_2018}
\bibfield{author}{\bibinfo{person}{Lian Loke} {and} \bibinfo{person}{Thecla Schiphorst}.} \bibinfo{year}{2018}\natexlab{}.
\newblock \showarticletitle{The somatic turn in human-computer interaction}.
\newblock \bibinfo{journal}{\emph{Interactions}} \bibinfo{volume}{25}, \bibinfo{number}{5} (\bibinfo{date}{Aug.} \bibinfo{year}{2018}), \bibinfo{pages}{54--5863}.
\newblock
\showISSN{1072-5520, 1558-3449}
\href{https://doi.org/10.1145/3236675}{doi:\nolinkurl{10.1145/3236675}}


\bibitem[Lygneg{\aa}rd et~al\mbox{.}(2023)]%
        {lygnegaard2023adolescent}
\bibfield{author}{\bibinfo{person}{Frida Lygneg{\aa}rd}, \bibinfo{person}{Maria Thell}, {and} \bibinfo{person}{Anna Sarkadi}.} \bibinfo{year}{2023}\natexlab{}.
\newblock \showarticletitle{Adolescent co-researchers identified the central role of social media for young people during the pandemic}.
\newblock \bibinfo{journal}{\emph{Acta Paediatrica}} \bibinfo{volume}{112}, \bibinfo{number}{4} (\bibinfo{year}{2023}), \bibinfo{pages}{787--793}.
\newblock


\bibitem[M{\'a}rquez~Segura et~al\mbox{.}(2016)]%
        {marquez2016embodied}
\bibfield{author}{\bibinfo{person}{Elena M{\'a}rquez~Segura}, \bibinfo{person}{Laia Turmo~Vidal}, \bibinfo{person}{Asreen Rostami}, {and} \bibinfo{person}{Annika Waern}.} \bibinfo{year}{2016}\natexlab{}.
\newblock \showarticletitle{Embodied sketching}. In \bibinfo{booktitle}{\emph{Proceedings of the 2016 CHI Conference on Human Factors in Computing Systems}}. \bibinfo{pages}{6014--6027}.
\newblock


\bibitem[McCabe and Ricciardelli(2003)]%
        {mccabe2003sociocultural}
\bibfield{author}{\bibinfo{person}{Marita~P McCabe} {and} \bibinfo{person}{Lina~A Ricciardelli}.} \bibinfo{year}{2003}\natexlab{}.
\newblock \showarticletitle{Sociocultural influences on body image and body changes among adolescent boys and girls}.
\newblock \bibinfo{journal}{\emph{The Journal of social psychology}} \bibinfo{volume}{143}, \bibinfo{number}{1} (\bibinfo{year}{2003}), \bibinfo{pages}{5--26}.
\newblock


\bibitem[McMahon et~al\mbox{.}(2020)]%
        {mcmahon2020stressful}
\bibfield{author}{\bibinfo{person}{Grace McMahon}, \bibinfo{person}{Ann-Marie Creaven}, {and} \bibinfo{person}{Stephen Gallagher}.} \bibinfo{year}{2020}\natexlab{}.
\newblock \showarticletitle{Stressful life events and adolescent well-being: The role of parent and peer relationships}.
\newblock \bibinfo{journal}{\emph{Stress and Health}} \bibinfo{volume}{36}, \bibinfo{number}{3} (\bibinfo{year}{2020}), \bibinfo{pages}{299--310}.
\newblock


\bibitem[Naseem et~al\mbox{.}(2020)]%
        {naseem_designing_2020}
\bibfield{author}{\bibinfo{person}{Mustafa Naseem}, \bibinfo{person}{Fouzia Younas}, {and} \bibinfo{person}{Maryam Mustafa}.} \bibinfo{year}{2020}\natexlab{}.
\newblock \showarticletitle{Designing {Digital} {Safe} {Spaces} for {Peer} {Support} and {Connectivity} in {Patriarchal} {Contexts}}.
\newblock \bibinfo{journal}{\emph{Proceedings of the ACM on Human-Computer Interaction}} \bibinfo{volume}{4}, \bibinfo{number}{CSCW2} (\bibinfo{date}{Oct.} \bibinfo{year}{2020}), \bibinfo{pages}{1--24}.
\newblock
\showISSN{2573-0142}
\href{https://doi.org/10.1145/3415217}{doi:\nolinkurl{10.1145/3415217}}


\bibitem[Núñez-Pacheco(2022)]%
        {nunez-pacheco_dialoguing_2022}
\bibfield{author}{\bibinfo{person}{Claudia Núñez-Pacheco}.} \bibinfo{year}{2022}\natexlab{}.
\newblock \showarticletitle{Dialoguing with {Tangible} {Traces}: {A} {Method} to {Elicit} {Autoethnographic} {Narratives}}. In \bibinfo{booktitle}{\emph{Sixteenth {International} {Conference} on {Tangible}, {Embedded}, and {Embodied} {Interaction}}}. \bibinfo{publisher}{ACM}, \bibinfo{address}{Daejeon Republic of Korea}, \bibinfo{pages}{1--14}.
\newblock
\showISBNx{978-1-4503-9147-4}
\href{https://doi.org/10.1145/3490149.3502255}{doi:\nolinkurl{10.1145/3490149.3502255}}


\bibitem[Organization(2024)]%
        {noauthor_mental_nodate}
\bibfield{author}{\bibinfo{person}{World~Health Organization}.} \bibinfo{year}{2024}\natexlab{}.
\newblock \bibinfo{title}{Mental health of adolescents}.
\newblock
\urldef\tempurl%
\url{https://www.who.int/news-room/fact-sheets/detail/adolescent-mental-health}
\showURL{%
\tempurl}
\newblock
\shownote{Accessed: 2024-09-28}.


\bibitem[Park and Lee(2020)]%
        {park_can_2020}
\bibfield{author}{\bibinfo{person}{Hyanghee Park} {and} \bibinfo{person}{Joonhwan Lee}.} \bibinfo{year}{2020}\natexlab{}.
\newblock \showarticletitle{Can a {Conversational} {Agent} {Lower} {Sexual} {Violence} {Victims}' {Burden} of {Self}-{Disclosure}?}. In \bibinfo{booktitle}{\emph{Extended {Abstracts} of the 2020 {CHI} {Conference} on {Human} {Factors} in {Computing} {Systems}}}. \bibinfo{publisher}{ACM}, \bibinfo{address}{Honolulu HI USA}.
\newblock
\href{https://doi.org/10.1145/3334480.3383050}{doi:\nolinkurl{10.1145/3334480.3383050}}


\bibitem[Peck et~al\mbox{.}(2020)]%
        {peck_inducing_2020}
\bibfield{author}{\bibinfo{person}{Tabitha~C. Peck}, \bibinfo{person}{Jessica~J. Good}, {and} \bibinfo{person}{Kimberly~A. Bourne}.} \bibinfo{year}{2020}\natexlab{}.
\newblock \showarticletitle{Inducing and {Mitigating} {Stereotype} {Threat} {Through} {Gendered} {Virtual} {Body}-{Swap} {Illusions}}. In \bibinfo{booktitle}{\emph{Proceedings of the 2020 {CHI} {Conference} on {Human} {Factors} in {Computing} {Systems}}}. \bibinfo{publisher}{ACM}, \bibinfo{address}{Honolulu HI USA}, \bibinfo{pages}{1--13}.
\newblock
\showISBNx{978-1-4503-6708-0}
\href{https://doi.org/10.1145/3313831.3376419}{doi:\nolinkurl{10.1145/3313831.3376419}}


\bibitem[Picard(2000)]%
        {picard2000affective}
\bibfield{author}{\bibinfo{person}{Rosalind~W Picard}.} \bibinfo{year}{2000}\natexlab{}.
\newblock \bibinfo{booktitle}{\emph{Affective computing}}.
\newblock \bibinfo{publisher}{MIT press}.
\newblock


\bibitem[Pinch et~al\mbox{.}(2024)]%
        {pinch_subtleties_2024}
\bibfield{author}{\bibinfo{person}{Annika Pinch}, \bibinfo{person}{Jeremy Birnholtz}, \bibinfo{person}{Kathryn Macapagal}, \bibinfo{person}{Ashley Kraus}, {and} \bibinfo{person}{David Moskowitz}.} \bibinfo{year}{2024}\natexlab{}.
\newblock \showarticletitle{The {Subtleties} of {Self}-{Presentation}: {A} study of sensitive disclosure among sexual minority adolescents}.
\newblock \bibinfo{journal}{\emph{Proceedings of the ACM on Human-Computer Interaction}} \bibinfo{volume}{8}, \bibinfo{number}{CSCW1} (\bibinfo{date}{April} \bibinfo{year}{2024}), \bibinfo{pages}{1--27}.
\newblock
\showISSN{2573-0142}
\href{https://doi.org/10.1145/3637408}{doi:\nolinkurl{10.1145/3637408}}
\newblock
\shownote{Publisher: Association for Computing Machinery (ACM)}.


\bibitem[Piran and Tylka(2019)]%
        {piran_handbook_2019}
\bibfield{editor}{\bibinfo{person}{Niva Piran} {and} \bibinfo{person}{Tracy~L. Tylka}} (Eds.). \bibinfo{year}{2019}\natexlab{}.
\newblock \bibinfo{booktitle}{\emph{Handbook of positive body image and embodiment: constructs, protective factors, and interventions}}.
\newblock \bibinfo{publisher}{Oxford University Press}, \bibinfo{address}{New York}.
\newblock
\showISBNx{978-0-19-084187-4}
\newblock
\shownote{OCLC: on1056781201}.


\bibitem[Price et~al\mbox{.}(2024)]%
        {price_childrens_2024}
\bibfield{author}{\bibinfo{person}{Linda Price}, \bibinfo{person}{Irum Rauf}, \bibinfo{person}{Daniel Gooch}, \bibinfo{person}{Dmitri Katz}, \bibinfo{person}{Oliver Pearce}, {and} \bibinfo{person}{Blaine Price}.} \bibinfo{year}{2024}\natexlab{}.
\newblock \showarticletitle{Children's perspectives on pain-logging: {Insights} from a {Co}-{Design} {Approach}}. In \bibinfo{booktitle}{\emph{Designing {Interactive} {Systems} {Conference}}}. \bibinfo{publisher}{ACM}, \bibinfo{address}{IT University of Copenhagen Denmark}, \bibinfo{pages}{1306--1318}.
\newblock
\showISBNx{9798400705830}
\href{https://doi.org/10.1145/3643834.3661597}{doi:\nolinkurl{10.1145/3643834.3661597}}


\bibitem[Rapo-Pylkk{\"o} et~al\mbox{.}(2016)]%
        {rapo2016chronic}
\bibfield{author}{\bibinfo{person}{Susanna Rapo-Pylkk{\"o}}, \bibinfo{person}{Maija Haanp{\"a}{\"a}}, {and} \bibinfo{person}{Helena Liira}.} \bibinfo{year}{2016}\natexlab{}.
\newblock \showarticletitle{Chronic pain among community-dwelling elderly: a population-based clinical study}.
\newblock \bibinfo{journal}{\emph{Scandinavian journal of primary health care}} \bibinfo{volume}{34}, \bibinfo{number}{2} (\bibinfo{year}{2016}), \bibinfo{pages}{159--164}.
\newblock


\bibitem[Rasouli et~al\mbox{.}(2022)]%
        {rasouli_proposed_2022}
\bibfield{author}{\bibinfo{person}{Samira Rasouli}, \bibinfo{person}{Garima Gupta}, \bibinfo{person}{Moojan Ghafurian}, {and} \bibinfo{person}{Kerstin Dautenhahn}.} \bibinfo{year}{2022}\natexlab{}.
\newblock \showarticletitle{Proposed {Applications} of {Social} {Robots} in {Interventions} for {Children} and {Adolescents} with {Social} {Anxiety}}. In \bibinfo{booktitle}{\emph{Sixteenth {International} {Conference} on {Tangible}, {Embedded}, and {Embodied} {Interaction}}}. \bibinfo{publisher}{ACM}, \bibinfo{address}{Daejeon Republic of Korea}, \bibinfo{pages}{1--7}.
\newblock
\showISBNx{978-1-4503-9147-4}
\href{https://doi.org/10.1145/3490149.3505575}{doi:\nolinkurl{10.1145/3490149.3505575}}


\bibitem[Ray et~al\mbox{.}(2013)]%
        {ray_use_2013}
\bibfield{author}{\bibinfo{person}{Dee~C. Ray}, \bibinfo{person}{Kasie~R. Lee}, \bibinfo{person}{Kristin~K. Meany-Walen}, \bibinfo{person}{Sarah~E. Carlson}, \bibinfo{person}{Kara~L. Carnes-Holt}, {and} \bibinfo{person}{Jenifer~N. Ware}.} \bibinfo{year}{2013}\natexlab{}.
\newblock \showarticletitle{Use of toys in child-centered play therapy.}
\newblock \bibinfo{journal}{\emph{International Journal of Play Therapy}} \bibinfo{volume}{22}, \bibinfo{number}{1} (\bibinfo{date}{Jan.} \bibinfo{year}{2013}), \bibinfo{pages}{43--57}.
\newblock
\showISSN{1939-0629, 1555-6824}
\href{https://doi.org/10.1037/a0031430}{doi:\nolinkurl{10.1037/a0031430}}


\bibitem[Reed et~al\mbox{.}(2023)]%
        {reed2023negotiating}
\bibfield{author}{\bibinfo{person}{Courtney~N Reed}, \bibinfo{person}{Paul Strohmeier}, {and} \bibinfo{person}{Andrew~P McPherson}.} \bibinfo{year}{2023}\natexlab{}.
\newblock \showarticletitle{Negotiating experience and communicating information through abstract metaphor}. In \bibinfo{booktitle}{\emph{Proceedings of the 2023 CHI Conference on Human Factors in Computing Systems}}. \bibinfo{pages}{1--16}.
\newblock


\bibitem[Sauer-Zavala et~al\mbox{.}(2013)]%
        {sauer2013comparing}
\bibfield{author}{\bibinfo{person}{Shannon~E Sauer-Zavala}, \bibinfo{person}{Erin~C Walsh}, \bibinfo{person}{Tory~A Eisenlohr-Moul}, {and} \bibinfo{person}{Emily~LB Lykins}.} \bibinfo{year}{2013}\natexlab{}.
\newblock \showarticletitle{Comparing mindfulness-based intervention strategies: Differential effects of sitting meditation, body scan, and mindful yoga}.
\newblock \bibinfo{journal}{\emph{Mindfulness}}  \bibinfo{volume}{4} (\bibinfo{year}{2013}), \bibinfo{pages}{383--388}.
\newblock


\bibitem[Schraml et~al\mbox{.}(2011a)]%
        {schraml2011stress}
\bibfield{author}{\bibinfo{person}{Karin Schraml}, \bibinfo{person}{Aleksander Perski}, \bibinfo{person}{Giorgio Grossi}, {and} \bibinfo{person}{Margareta Simonsson-Sarnecki}.} \bibinfo{year}{2011}\natexlab{a}.
\newblock \showarticletitle{Stress symptoms among adolescents: The role of subjective psychosocial conditions, lifestyle, and self-esteem}.
\newblock \bibinfo{journal}{\emph{Journal of adolescence}} \bibinfo{volume}{34}, \bibinfo{number}{5} (\bibinfo{year}{2011}), \bibinfo{pages}{987--996}.
\newblock


\bibitem[Schraml et~al\mbox{.}(2011b)]%
        {schraml_stress_2011}
\bibfield{author}{\bibinfo{person}{Karin Schraml}, \bibinfo{person}{Aleksander Perski}, \bibinfo{person}{Giorgio Grossi}, {and} \bibinfo{person}{Margareta Simonsson‐Sarnecki}.} \bibinfo{year}{2011}\natexlab{b}.
\newblock \showarticletitle{Stress symptoms among adolescents: {The} role of subjective psychosocial conditions, lifestyle, and self‐esteem}.
\newblock \bibinfo{journal}{\emph{Journal of Adolescence}} \bibinfo{volume}{34}, \bibinfo{number}{5} (\bibinfo{date}{Oct.} \bibinfo{year}{2011}), \bibinfo{pages}{987--996}.
\newblock
\showISSN{0140-1971, 1095-9254}
\href{https://doi.org/10.1016/j.adolescence.2010.11.010}{doi:\nolinkurl{10.1016/j.adolescence.2010.11.010}}


\bibitem[Shi et~al\mbox{.}(2024)]%
        {shi_disandbox_2024}
\bibfield{author}{\bibinfo{person}{Yan Shi}, \bibinfo{person}{Lidan Gong}, \bibinfo{person}{Yiwen Lu}, {and} \bibinfo{person}{Lijuan Liu}.} \bibinfo{year}{2024}\natexlab{}.
\newblock \showarticletitle{{DiSandbox}: {A} {Low}-cost {Digital} {Sandbox} {Tool} to {Support} {Psychological} {Analysis} and {Therapy} for {Left}-behind {Children}}. In \bibinfo{booktitle}{\emph{Extended {Abstracts} of the {CHI} {Conference} on {Human} {Factors} in {Computing} {Systems}}}. \bibinfo{publisher}{ACM}, \bibinfo{address}{Honolulu HI USA}, \bibinfo{pages}{1--6}.
\newblock
\showISBNx{9798400703317}
\href{https://doi.org/10.1145/3613905.3650906}{doi:\nolinkurl{10.1145/3613905.3650906}}


\bibitem[Sinton and Birch(2006)]%
        {sinton2006individual}
\bibfield{author}{\bibinfo{person}{Meghan~M Sinton} {and} \bibinfo{person}{Leann~L Birch}.} \bibinfo{year}{2006}\natexlab{}.
\newblock \showarticletitle{Individual and sociocultural influences on pre-adolescent girls’ appearance schemas and body dissatisfaction}.
\newblock \bibinfo{journal}{\emph{Journal of Youth and Adolescence}}  \bibinfo{volume}{35} (\bibinfo{year}{2006}), \bibinfo{pages}{157--167}.
\newblock


\bibitem[Soleymani et~al\mbox{.}(2019)]%
        {soleymani_multimodal_2019}
\bibfield{author}{\bibinfo{person}{Mohammad Soleymani}, \bibinfo{person}{Kalin Stefanov}, \bibinfo{person}{Sin-Hwa Kang}, \bibinfo{person}{Jan Ondras}, {and} \bibinfo{person}{Jonathan Gratch}.} \bibinfo{year}{2019}\natexlab{}.
\newblock \showarticletitle{Multimodal {Analysis} and {Estimation} of {Intimate} {Self}-{Disclosure}}. In \bibinfo{booktitle}{\emph{2019 {International} {Conference} on {Multimodal} {Interaction}}}. \bibinfo{publisher}{ACM}, \bibinfo{address}{Suzhou China}, \bibinfo{pages}{59--68}.
\newblock
\showISBNx{978-1-4503-6860-5}
\href{https://doi.org/10.1145/3340555.3353737}{doi:\nolinkurl{10.1145/3340555.3353737}}


\bibitem[Sontag et~al\mbox{.}(2008)]%
        {sontag_coping_2008}
\bibfield{author}{\bibinfo{person}{Lisa~M. Sontag}, \bibinfo{person}{Julia~A. Graber}, \bibinfo{person}{Jeanne Brooks-Gunn}, {and} \bibinfo{person}{Michelle~P. Warren}.} \bibinfo{year}{2008}\natexlab{}.
\newblock \showarticletitle{Coping with {Social} {Stress}: {Implications} for {Psychopathology} in {Young} {Adolescent} {Girls}}.
\newblock \bibinfo{journal}{\emph{Journal of Abnormal Child Psychology}} \bibinfo{volume}{36}, \bibinfo{number}{8} (\bibinfo{date}{Nov.} \bibinfo{year}{2008}), \bibinfo{pages}{1159--1174}.
\newblock
\showISSN{0091-0627, 1573-2835}
\href{https://doi.org/10.1007/s10802-008-9239-3}{doi:\nolinkurl{10.1007/s10802-008-9239-3}}


\bibitem[Sun et~al\mbox{.}(2012)]%
        {sun2012academic}
\bibfield{author}{\bibinfo{person}{Jiandong Sun}, \bibinfo{person}{Michael~P Dunne}, {and} \bibinfo{person}{Xiang-Yu Hou}.} \bibinfo{year}{2012}\natexlab{}.
\newblock \showarticletitle{Academic stress among adolescents in China}.
\newblock \bibinfo{journal}{\emph{Australasian Epidemiologist}} \bibinfo{volume}{19}, \bibinfo{number}{1} (\bibinfo{year}{2012}), \bibinfo{pages}{9--12}.
\newblock


\bibitem[Sun et~al\mbox{.}(2013)]%
        {sun2013educational}
\bibfield{author}{\bibinfo{person}{Jiandong Sun}, \bibinfo{person}{Michael~P Dunne}, \bibinfo{person}{Xiang-yu Hou}, {and} \bibinfo{person}{Ai-qiang Xu}.} \bibinfo{year}{2013}\natexlab{}.
\newblock \showarticletitle{Educational stress among Chinese adolescents: Individual, family, school and peer influences}.
\newblock \bibinfo{journal}{\emph{Educational Review}} \bibinfo{volume}{65}, \bibinfo{number}{3} (\bibinfo{year}{2013}), \bibinfo{pages}{284--302}.
\newblock


\bibitem[Susman et~al\mbox{.}(1988)]%
        {susman1988physiological}
\bibfield{author}{\bibinfo{person}{Elizabeth~J Susman}, \bibinfo{person}{Editha~D Nottelmann}, \bibinfo{person}{Lorah~D Dorn}, \bibinfo{person}{Gale Inoff-Germain}, {and} \bibinfo{person}{George~P Chrousos}.} \bibinfo{year}{1988}\natexlab{}.
\newblock \showarticletitle{Physiological and behavioral aspects of stress in adolescence}.
\newblock \bibinfo{journal}{\emph{Mechanisms of physical and emotional stress}} (\bibinfo{year}{1988}), \bibinfo{pages}{341--352}.
\newblock


\bibitem[Søndergaard et~al\mbox{.}(2021a)]%
        {sondergaard_designing_2021}
\bibfield{author}{\bibinfo{person}{Marie Louise~Juul Søndergaard}, \bibinfo{person}{Marianela Ciolfi~Felice}, {and} \bibinfo{person}{Madeline Balaam}.} \bibinfo{year}{2021}\natexlab{a}.
\newblock \showarticletitle{Designing {Menstrual} {Technologies} with {Adolescents}}. In \bibinfo{booktitle}{\emph{Proceedings of the 2021 {CHI} {Conference} on {Human} {Factors} in {Computing} {Systems}}}. \bibinfo{publisher}{ACM}, \bibinfo{address}{Yokohama Japan}, \bibinfo{pages}{1--14}.
\newblock
\showISBNx{978-1-4503-8096-6}
\href{https://doi.org/10.1145/3411764.3445471}{doi:\nolinkurl{10.1145/3411764.3445471}}


\bibitem[Søndergaard et~al\mbox{.}(2021b)]%
        {sondergaard_designing_2021-1}
\bibfield{author}{\bibinfo{person}{Marie Louise~Juul Søndergaard}, \bibinfo{person}{Marianela Ciolfi~Felice}, {and} \bibinfo{person}{Madeline Balaam}.} \bibinfo{year}{2021}\natexlab{b}.
\newblock \showarticletitle{Designing {Menstrual} {Technologies} with {Adolescents}}. In \bibinfo{booktitle}{\emph{Proceedings of the 2021 {CHI} {Conference} on {Human} {Factors} in {Computing} {Systems}}}. \bibinfo{publisher}{ACM}, \bibinfo{address}{Yokohama Japan}, \bibinfo{pages}{1--14}.
\newblock
\showISBNx{978-1-4503-8096-6}
\href{https://doi.org/10.1145/3411764.3445471}{doi:\nolinkurl{10.1145/3411764.3445471}}


\bibitem[Søndergaard et~al\mbox{.}(2020)]%
        {sondergaard_designing_2020}
\bibfield{author}{\bibinfo{person}{Marie Louise~Juul Søndergaard}, \bibinfo{person}{Ozgun Kilic~Afsar}, \bibinfo{person}{Marianela Ciolfi~Felice}, \bibinfo{person}{Nadia Campo~Woytuk}, {and} \bibinfo{person}{Madeline Balaam}.} \bibinfo{year}{2020}\natexlab{}.
\newblock \showarticletitle{Designing with {Intimate} {Materials} and {Movements}: {Making} "{Menarche} {Bits}"}. In \bibinfo{booktitle}{\emph{Proceedings of the 2020 {ACM} {Designing} {Interactive} {Systems} {Conference}}}. \bibinfo{publisher}{ACM}, \bibinfo{address}{Eindhoven Netherlands}, \bibinfo{pages}{587--600}.
\newblock
\showISBNx{978-1-4503-6974-9}
\href{https://doi.org/10.1145/3357236.3395592}{doi:\nolinkurl{10.1145/3357236.3395592}}


\bibitem[Tate et~al\mbox{.}(2022)]%
        {tate2022personality}
\bibfield{author}{\bibinfo{person}{Christopher Tate}, \bibinfo{person}{Rajnish Kumar}, \bibinfo{person}{Jennifer~M Murray}, \bibinfo{person}{Sharon Sanchez-Franco}, \bibinfo{person}{Olga~L Sarmiento}, \bibinfo{person}{Shannon~C Montgomery}, \bibinfo{person}{Huiyu Zhou}, \bibinfo{person}{Abhijit Ramalingam}, \bibinfo{person}{Erin Krupka}, \bibinfo{person}{Erik Kimbrough}, {et~al\mbox{.}}} \bibinfo{year}{2022}\natexlab{}.
\newblock \showarticletitle{The personality and cognitive traits associated with adolescents’ sensitivity to social norms}.
\newblock \bibinfo{journal}{\emph{Scientific Reports}} \bibinfo{volume}{12}, \bibinfo{number}{1} (\bibinfo{year}{2022}), \bibinfo{pages}{15247}.
\newblock


\bibitem[Tsui and Rich(2002)]%
        {tsui2002only}
\bibfield{author}{\bibinfo{person}{Ming Tsui} {and} \bibinfo{person}{Lynne Rich}.} \bibinfo{year}{2002}\natexlab{}.
\newblock \showarticletitle{The only child and educational opportunity for girls in urban China}.
\newblock \bibinfo{journal}{\emph{Gender \& Society}} \bibinfo{volume}{16}, \bibinfo{number}{1} (\bibinfo{year}{2002}), \bibinfo{pages}{74--92}.
\newblock


\bibitem[Umair et~al\mbox{.}(2020)]%
        {umair_thermopixels_2020}
\bibfield{author}{\bibinfo{person}{Muhammad Umair}, \bibinfo{person}{Corina Sas}, {and} \bibinfo{person}{Miquel Alfaras}.} \bibinfo{year}{2020}\natexlab{}.
\newblock \showarticletitle{{ThermoPixels}: {Toolkit} for {Personalizing} {Arousal}-based {Interfaces} through {Hybrid} {Crafting}}. In \bibinfo{booktitle}{\emph{Proceedings of the 2020 {ACM} {Designing} {Interactive} {Systems} {Conference}}}. \bibinfo{publisher}{ACM}, \bibinfo{address}{Eindhoven Netherlands}, \bibinfo{pages}{1017--1032}.
\newblock
\showISBNx{978-1-4503-6974-9}
\href{https://doi.org/10.1145/3357236.3395512}{doi:\nolinkurl{10.1145/3357236.3395512}}


\bibitem[Velthouse(1990)]%
        {velthouse1990creativity}
\bibfield{author}{\bibinfo{person}{Betty~A Velthouse}.} \bibinfo{year}{1990}\natexlab{}.
\newblock \showarticletitle{Creativity and empowerment: a complementary relationship}.
\newblock \bibinfo{journal}{\emph{Review of Business}} \bibinfo{volume}{12}, \bibinfo{number}{2} (\bibinfo{year}{1990}), \bibinfo{pages}{13}.
\newblock


\bibitem[Vitetta et~al\mbox{.}(2005)]%
        {vitetta_mindbody_2005}
\bibfield{author}{\bibinfo{person}{L Vitetta}, \bibinfo{person}{B Anton}, \bibinfo{person}{F Cortizo}, {and} \bibinfo{person}{A Sali}.} \bibinfo{year}{2005}\natexlab{}.
\newblock \showarticletitle{Mind‐{Body} {Medicine}: {Stress} and {Its} {Impact} on {Overall} {Health} and {Longevity}}.
\newblock \bibinfo{journal}{\emph{Annals of the New York Academy of Sciences}} \bibinfo{volume}{1057}, \bibinfo{number}{1} (\bibinfo{date}{Dec.} \bibinfo{year}{2005}), \bibinfo{pages}{492--505}.
\newblock
\showISSN{0077-8923, 1749-6632}
\href{https://doi.org/10.1111/j.1749-6632.2005.tb06153.x}{doi:\nolinkurl{10.1111/j.1749-6632.2005.tb06153.x}}


\bibitem[Waldie(2001)]%
        {waldie2001childhood}
\bibfield{author}{\bibinfo{person}{Karen~E Waldie}.} \bibinfo{year}{2001}\natexlab{}.
\newblock \showarticletitle{Childhood headache, stress in adolescence, and primary headache in young adulthood: a longitudinal cohort study}.
\newblock \bibinfo{journal}{\emph{Headache: The Journal of Head and Face Pain}} \bibinfo{volume}{41}, \bibinfo{number}{1} (\bibinfo{year}{2001}), \bibinfo{pages}{1--10}.
\newblock


\bibitem[Wallace et~al\mbox{.}(2013)]%
        {wallace_making_2013}
\bibfield{author}{\bibinfo{person}{Jayne Wallace}, \bibinfo{person}{John McCarthy}, \bibinfo{person}{Peter~C. Wright}, {and} \bibinfo{person}{Patrick Olivier}.} \bibinfo{year}{2013}\natexlab{}.
\newblock \showarticletitle{Making design probes work}. In \bibinfo{booktitle}{\emph{Proceedings of the {SIGCHI} {Conference} on {Human} {Factors} in {Computing} {Systems}}}. \bibinfo{publisher}{ACM}, \bibinfo{address}{Paris France}, \bibinfo{pages}{3441--3450}.
\newblock
\showISBNx{978-1-4503-1899-0}
\href{https://doi.org/10.1145/2470654.2466473}{doi:\nolinkurl{10.1145/2470654.2466473}}


\bibitem[Wang et~al\mbox{.}(2023b)]%
        {wang_emtex_2023}
\bibfield{author}{\bibinfo{person}{Qi Wang}, \bibinfo{person}{Yuan Zeng}, \bibinfo{person}{Runhua Zhang}, \bibinfo{person}{Nianding Ye}, \bibinfo{person}{Linghao Zhu}, \bibinfo{person}{Xiaohua Sun}, {and} \bibinfo{person}{Teng Han}.} \bibinfo{year}{2023}\natexlab{b}.
\newblock \showarticletitle{{EmTex}: {Prototyping} {Textile}-{Based} {Interfaces} through {An} {Embroidered} {Construction} {Kit}}. In \bibinfo{booktitle}{\emph{Proceedings of the 36th {Annual} {ACM} {Symposium} on {User} {Interface} {Software} and {Technology}}}. \bibinfo{publisher}{ACM}, \bibinfo{address}{San Francisco CA USA}, \bibinfo{pages}{1--17}.
\newblock
\showISBNx{9798400701320}
\href{https://doi.org/10.1145/3586183.3606815}{doi:\nolinkurl{10.1145/3586183.3606815}}


\bibitem[Wang et~al\mbox{.}(2023a)]%
        {wang_large-scale_2023}
\bibfield{author}{\bibinfo{person}{Xin Wang}, \bibinfo{person}{Yan Wang}, \bibinfo{person}{Xiyan Zhang}, \bibinfo{person}{Wenyi Yang}, \bibinfo{person}{Jie Yang}, {and} \bibinfo{person}{{Department of Child and Adolescent Health Promotion, Jiangsu Provincial Center for Disease Control and Prevention, Nanjing City, Jiangsu Province, China}}.} \bibinfo{year}{2023}\natexlab{a}.
\newblock \showarticletitle{A {Large}-{Scale} {Cross}-{Sectional} {Study} on {Mental} {Health} {Status} {Among} {Children} and {Adolescents} — {Jiangsu} {Province}, {China}, 2022}.
\newblock \bibinfo{journal}{\emph{China CDC Weekly}} \bibinfo{volume}{5}, \bibinfo{number}{32} (\bibinfo{year}{2023}), \bibinfo{pages}{710--714}.
\newblock
\showISSN{2097-3101}
\href{https://doi.org/10.46234/ccdcw2023.136}{doi:\nolinkurl{10.46234/ccdcw2023.136}}


\bibitem[Willander and Larsson(2006)]%
        {willander_smell_2006}
\bibfield{author}{\bibinfo{person}{Johan Willander} {and} \bibinfo{person}{Maria Larsson}.} \bibinfo{year}{2006}\natexlab{}.
\newblock \showarticletitle{Smell your way back to childhood: {Autobiographical} odor memory}.
\newblock \bibinfo{journal}{\emph{Psychonomic Bulletin \& Review}} \bibinfo{volume}{13}, \bibinfo{number}{2} (\bibinfo{date}{April} \bibinfo{year}{2006}), \bibinfo{pages}{240--244}.
\newblock
\showISSN{1069-9384, 1531-5320}
\href{https://doi.org/10.3758/BF03193837}{doi:\nolinkurl{10.3758/BF03193837}}


\bibitem[Yaribeygi et~al\mbox{.}(2017)]%
        {yaribeygi2017impact}
\bibfield{author}{\bibinfo{person}{Habib Yaribeygi}, \bibinfo{person}{Yunes Panahi}, \bibinfo{person}{Hedayat Sahraei}, \bibinfo{person}{Thomas~P Johnston}, {and} \bibinfo{person}{Amirhossein Sahebkar}.} \bibinfo{year}{2017}\natexlab{}.
\newblock \showarticletitle{The impact of stress on body function: A review}.
\newblock \bibinfo{journal}{\emph{EXCLI journal}}  \bibinfo{volume}{16} (\bibinfo{year}{2017}), \bibinfo{pages}{1057}.
\newblock


\bibitem[Yu et~al\mbox{.}(2018)]%
        {yu_biofeedback_2018}
\bibfield{author}{\bibinfo{person}{Bin Yu}, \bibinfo{person}{Mathias Funk}, \bibinfo{person}{Jun Hu}, \bibinfo{person}{Qi Wang}, {and} \bibinfo{person}{Loe Feijs}.} \bibinfo{year}{2018}\natexlab{}.
\newblock \showarticletitle{Biofeedback for {Everyday} {Stress} {Management}: {A} {Systematic} {Review}}.
\newblock \bibinfo{journal}{\emph{Frontiers in ICT}}  \bibinfo{volume}{5} (\bibinfo{date}{Sept.} \bibinfo{year}{2018}), \bibinfo{pages}{23}.
\newblock
\showISSN{2297-198X}
\href{https://doi.org/10.3389/fict.2018.00023}{doi:\nolinkurl{10.3389/fict.2018.00023}}


\end{thebibliography}

\clearpage
\appendix
\balance 
\section{Participants’ Design Responses}
In this appendix we present participants’ design responses by the end of Embodied Probes workshop. These designs were informed by the stress-related insights and stories shared by all the participants through different embodied methods at the workshop. The ideas are presented using the participants’ own words to authentically capture their perspectives.

\textbf{\textit{Participant 1,2 and 3 created “Hell Girl”, }}a piece of stress-relieving toy (\autoref{fig7}, left). Drawing from the cultural reference of Voodoo dolls and western horror films, the toy features the visual representations of a scarecrow and functions of a Voodoo doll. P2 explained: \textit{“We created a tool for venting emotions—a scarecrow. The scarecrow is equipped with soft pressure sensor, and pressing it makes it screams, allowing you to imagine the scarecrow as someone you dislike. By squeezing it to generate sounds, you can release your stress.”} P3 emphasised the emotional benefit of the design : \textit{“It feels like you have the person you dislike completely under your control, which is incredibly satisfying.”} The intended use context of this design is after school, when the user is alone by herself. The sounds triggered by squeezing the scarecrow could also be customised by the user to be evocative.

\textbf{\textit{Participant 4,5 and 6 created “Riffing T-shirt”, }} a T-shirt that could be played as a guitar (\autoref{fig7}, middle). P5 explained their insights towards stress-relief: \textit{“We designed this device to help release stress. We noticed that when people feel stressed, they often turn to auditory experiences for comfort and healing. So, we decided to combine auditory elements with clothing to create an integrated solution.”} P6 further explained the interaction design: \textit{“This piece of clothing is a musical instrument. From the beginning, we wanted the instrument on the clothing to be playable, so I designed it in the style of a guitar. When someone plucks the guitar strings, the stretch sensors activate and play music, helping to relieve stress.”} The group also designed two additional T-shirts that share a similar concept and interactive design. These T-shirts are intended to be worn together with friends.

\textbf{\textit{Participant 7,8 and 9 created “EmoMic”, }}  an interactive microphone for stress-relief (\autoref{fig7}, right). This concept was informed by P8’s personal stress management strategy: \textit{“When people are unhappy, they may suddenly start singing at home to vent their emotions. Inspired by this, we came up with the idea of creating a microphone.”} The interaction design of the microphone is stated as follow: \textit{“We embedded a pressure sensor into the microphone: the harder you press, the higher the pitch it produces, resembling the way Minions speak. Conversely, a lighter press creates a much deeper tone, like a low ‘uncle voice’. We believe these playful sound effects can brighten your mood, and the altered voice adds a sense of comfort, like sharing your feelings with someone.”} The design is intended for use either with close friends or individually by the user. The design of “EmoMic” was also a fusion of P7 and P9’s creative ideas as described next.

\textbf{\textit{“Sentient Pen” by P7: }}  P7 sketched her idea for a pen that changes color based on the user's mood, envisioning the design as follows: \textit{“The Sentient Pen is a color-changing pen that alters its main color tone based on your emotions. The main color tone includes a wide range of colors, and when you use this pen to draw, the overall color scheme of your artwork will shift according to your emotional state. ”} P7's idea could not be turned into a working prototype but informed the group’s concept for Emomic.

\textbf{\textit{“MoodMirror Abebe” by P9:}}  P9 explained that her idea for creating a plush toy was inspired by a motivation to combine textiles with emotional management. \textit{“There is a pressure sensor embedded in the plush toy, which can select different modes by different strengths of pressing, there is a colour changing mode and a comforting mode: the colour changing mode simply expresses your emotions; different colours show different emotions, and the expression mode can comfort you, such as having a conversation with you or telling you jokes.”} Due to limited time and resources, P9's idea could not be developed into a working prototype during the workshop. As a result, the team decided to incorporate a dog-like plush toy onto the EmoMic, both to honor P9's concept and to add an additional layer of tactile interaction to the microphone.

\begin{figure*}[h]
    \centering
    \includegraphics[width=1\linewidth]{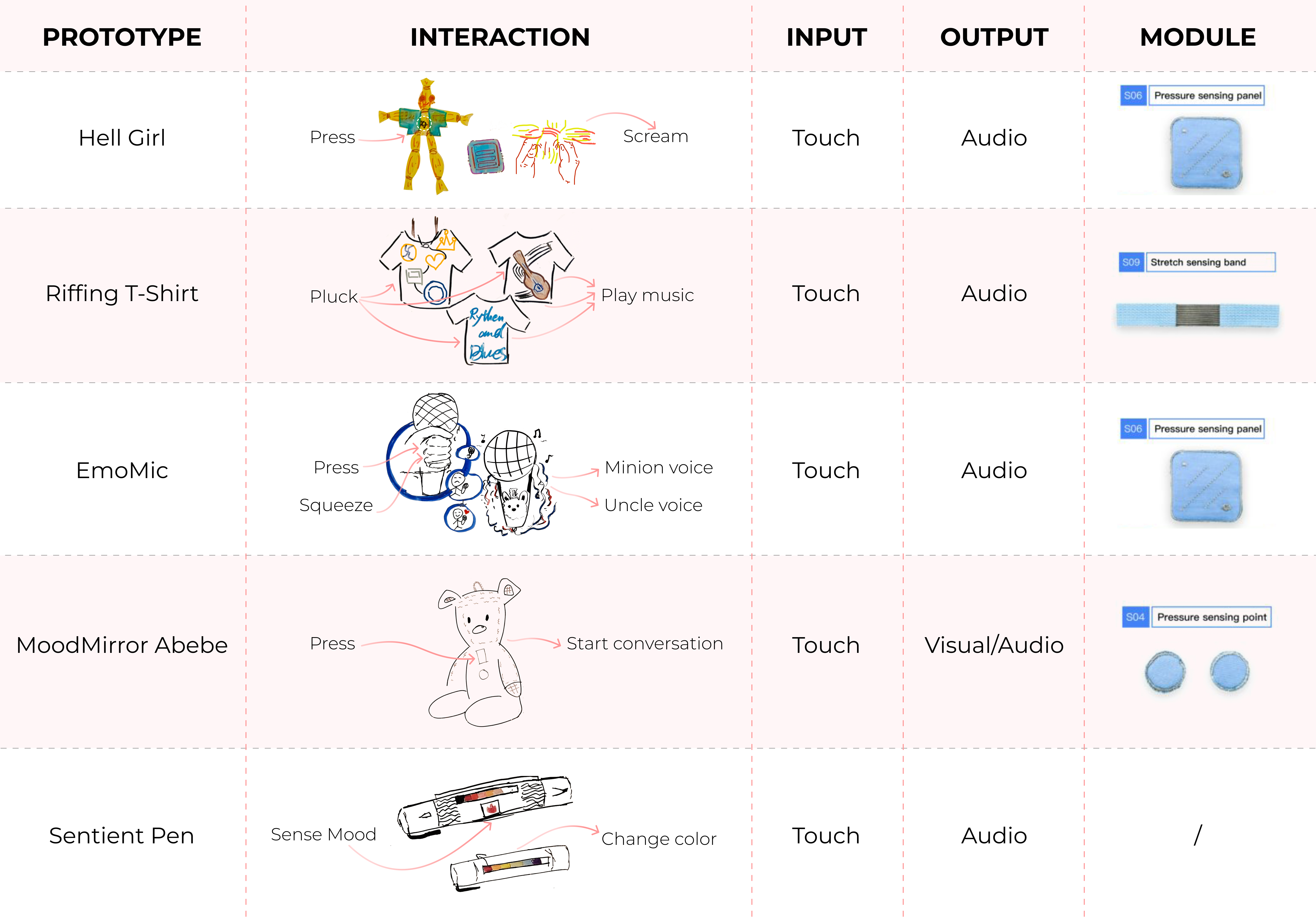}
    \caption{Interaction design of the prototypes created by participants. (All the illustrations in “Interaction” were created by the participants, photo credit for “Module”\cite{wang_emtex_2023})}
    \Description{Fig. 7: A table displaying five creative prototypes—Hell Girl, Riffing T-Shirt, EmoMic, MoodMirror Abebe, and Sentient Pen—each with illustrated interaction mechanisms. Columns detail how users engage with each prototype (e.g., press, pluck, squeeze), the type of input (Touch), output (Audio or Visual/Audio), and the sensor module used (e.g., pressure sensing panel, stretch sensing band). Illustrations visually depict the user actions and resulting effects, such as a toy screaming, music playing from a shirt, or a pen changing colour.}
    \label{fig8}
\end{figure*}

We analysed the five designs created by the participants to identify common input and output modalities for stress relief, which might inform future designs and technologies aimed at supporting adolescents’ mental well-being (\autoref{fig8}). Vision, as the dominant modality, was only used twice in the design concepts: “Sentient Pen” and “MoodMirror Abebe” both utilised vision as an output modality. Four designs (“Hell Girl”, “Riffing T-shirt”, “EmoMic”, and “MoodMirror Abebe”) employed audio as the output modality and touch as the input modality. The three final concepts, developed into working prototypes by three different groups, all used touch as the input modality and audio as the output modality. Participants reported that the sound feedback generated by touch, especially during actions like squeezing, was the most stress-relieving. The design of “MoodMirror Abebe” linked the strength of the user's touch to their stress level, while “EmoMic” connected the strength of squeezing to the pitch of the sound produced by the microphone, reflecting the user's emotional state. 

\end{document}